\PassOptionsToPackage{prologue,dvipsnames,svgnames}{xcolor}
\PassOptionsToPackage{hyphens}{url}
\documentclass[11pt]{article}

\usepackage[T1]{fontenc}
\usepackage{lmodern}

\usepackage{comment}
\usepackage{amsmath,amssymb,amsfonts,amsthm,mathtools}
\usepackage{mdframed}
\usepackage{todonotes}
\usepackage{subcaption}
\usepackage{titlesec}
\usepackage{booktabs}
\usepackage[ruled]{algorithm2e}
\usepackage{tikz}
\usetikzlibrary{positioning,patterns,decorations.pathreplacing,arrows.meta,shapes.geometric,calc,matrix,fit}
\usepackage{geometry}
\usepackage{nicefrac}
\usepackage{paralist}
\usepackage{enumitem}
\usepackage{xcolor}
\usepackage{colortbl}
\usepackage{listings}
\usepackage{fontawesome}
\usepackage{multirow}
\usepackage{tabularx}
\usepackage{pgfplots}
\pgfplotsset{compat=1.18}
\usepackage{pifont}
\usepackage{thmtools,thm-restate}
\usepackage{setspace}
\usepackage{etoolbox}
\usepackage{tablefootnote}
\usepackage{tocloft}
\usepackage{multicol}
\usepackage[authoryear,round]{natbib}
\usepackage{hyperref}
\usepackage{url}
\usepackage[sort&compress,nameinlink]{cleveref}
\usepackage{xcolor}

\definecolor{winered}{RGB}{114,47,55}

\let\oldemph\emph
\renewcommand{\emph}[1]{\textcolor{winered}{\oldemph{#1}}}
\hypersetup{
  colorlinks=true,
  urlcolor=blue,
  linkcolor=DarkBlue,
  citecolor=DarkGreen
}

\SetAlCapFnt{\small}
\SetAlCapNameFnt{\small}
\SetAlCapHSkip{0pt}
\IncMargin{-\parindent}

\title{}
\author{}
\date{}

\definecolor{winered}{rgb}{0.5,0.1,0.1}
\definecolor{darkgreen}{rgb}{0,0.35,0}
\definecolor{verylightgray}{gray}{0.9}
\definecolor{crimson}{RGB}{220,20,60}

\newcolumntype{Y}{>{\centering\arraybackslash}X}
\newcolumntype{Z}{>{\raggedleft\arraybackslash}X}
\newcolumntype{a}{>{\columncolor{black!05}}Y}

\newcommand{\one}{\mathbf{1}}
\renewcommand*{\le}{\leqslant}
\renewcommand*{\leq}{\leqslant}
\renewcommand*{\ge}{\geqslant}
\renewcommand*{\geq}{\geqslant}
\renewcommand{\epsilon}{\varepsilon}

\newcommand{\R}{\mathbb{R}}

\renewcommand{\part}{\mathrm{part}}
\renewcommand{\H}{\mathrm{H}}

\newcommand{\deferredsym}{\(\blacklozenge\)}

\DeclareMathOperator{\sgn}{sgn}

\newcommand{\inner}[2]{\langle #1,#2 \rangle}
\newcommand{\totalloss}{\mathcal{L}}

\newcommand{\lv}[1]{}

\newcommand{\eps}{\varepsilon}

\DeclareMathOperator*{\argmax}{arg\,max}
\DeclareMathOperator*{\argmin}{arg\,min}

\allowdisplaybreaks
\renewcommand{\eqref}[1]{(\ref{#1})}

\Crefname{table}{Table}{Tables}
\Crefname{figure}{Figure}{Figures}
\Crefname{theorem}{Theorem}{Theorems}
\Crefname{definition}{Definition}{Definitions}
\Crefname{corollary}{Corollary}{Corollaries}
\Crefname{observation}{Observation}{Observations}
\Crefname{question}{Question}{Question}
\Crefname{lemma}{Lemma}{Lemmas}
\Crefname{example}{Example}{Examples}
\Crefname{reduction}{Reduction}{Reductions}
\Crefname{construction}{Construction}{Constructions}
\Crefname{subsection}{Section}{Sections}
\Crefname{section}{Section}{Sections}
\Crefname{proposition}{Proposition}{Propositions}
\Crefname{algorithm}{Algorithm}{Algorithms}
\Crefname{algocf}{Algorithm}{Algorithms}
\Crefname{lstlisting}{listing}{listings}
\Crefname{equation}{Relation}{Relations}
\Crefname{claim}{Claim}{Claims}
\Crefname{fact}{Fact}{Facts}

\newtheoremstyle{ex}
{}{}{}{}{\bfseries}{.}{ }
{\thmname{#1}\thmnumber{ #2}\thmnote{{\bfseries: (#3)}}}

\theoremstyle{definition}
\newtheorem{definition}{Definition}

\newtheorem{fact}{Fact}

\theoremstyle{plain}
\newtheorem{theorem}{Theorem}

\newtheorem{claim}{Claim}

\newtheorem{lemma}{Lemma}

\begin{document}

\newgeometry{left=0.95in, right=0.95in, top=0.5in, bottom=0.8in}

\begin{flushleft}
{\fontsize{16.8}{18.5}\selectfont\bfseries
\scalebox{0.97}[1]{Enforcing Axioms for AI Alignment under Loss-Based Rules}\par}
\end{flushleft}

\vspace{0.5em}

\begin{center}
\large
\begin{tabular}{c@{\hspace{2cm}}c}
ALEXANDROS HOLLENDER\textsuperscript{*}&
SONJA KRAICZY\textsuperscript{*}$^\dagger$ \\
{\small Technical University of Munich\footnotemark[1]} &
{\small University of Oxford, Principles of Intelligence}
\end{tabular}

\vspace{0.4em}

{\footnotesize
\textsuperscript{*}Authors are listed in alphabetical order.\par
$^\dagger$ Correspondence to: \texttt{sonja.kraiczy@cs.ox.ac.uk}
}
\end{center}
\footnotetext[1]{Work done while at the University of Oxford.}
    {\small\noindent

Many alignment methods, most notably reinforcement learning from human feedback (RLHF), train reward models by minimizing a loss on pairwise comparisons of model responses. Inspired by principle-guided methods such as Constitutional AI, we treat each principle as a voter that compares responses according to that principle—for example, preferring the more helpful or less toxic response. We represent these voters as linear directions in a fixed embedding space, a simplifying assumption motivated by the Linear Representation Hypothesis.

In this \emph{linear social choice model}, Ge et al.~(2024) showed that an optimal linear reward model can violate Pareto optimality (PO): From the principles-as-voters lens, this means a response A can be less helpful and more toxic than B, yet still receive a higher reward. 
We ask whether richer reward models or changes to comparison data can recover PO within loss-based training. We show that PO can still fail for polynomials of any fixed degree and one-hidden-layer ReLU multilayer perceptrons (MLPs) of any fixed width \(d\ge2\) with suitable finite parameter bounds.
We then recover PO for loss-based training of linear rewards under suitable assumptions in an idealized setting with uniform comparison directions and equal comparison lengths.  These results motivate \emph{dataset design} as an approach to recovering PO in constitutional-style alignment while retaining loss-based training.
}

\setcounter{tocdepth}{2}

\vspace{0.8cm}
\noindent\rule{\textwidth}{0.8pt}

\begingroup
\renewcommand{\contentsname}{}
\begin{center}
    \textsc{Contents}
\end{center}
\vspace{-1.2cm}
\begin{spacing}{0.9}
    \tableofcontents
\end{spacing}
\endgroup

\noindent\rule{\textwidth}{0.5pt}

\setcounter{secnumdepth}{2}
\newgeometry{left=0.95in, right=0.95in, top=0.75in, bottom=0.8in}

\section{Introduction}
Many recent alignment methods follow a common pipeline: collect pairwise (binary) preferences over model responses, fit a reward (or preference) model to these comparisons, and then optimize the base model to minimize a loss correlated with aligned behavior. 
In the classical human-in-the-loop setting of Reinforcement Learning from Human Feedback (RLHF) \citep{christiano2017deep}, human annotators select the preferred response between candidate responses. Rather than eliciting unguided preferences, Anthropic introduced principle-oriented feedback \citep{bai2022training} in which annotators judge responses against explicit principles—helpfulness, honesty, and harmlessness (HHH); this principle-guided supervision later informed Constitutional AI, which formalizes a written set of principles to guide binary comparisons of model responses \citep{bai2022constitutional}. We use “constitutional-style” as an umbrella term for such principle-guided supervision.

Adopting a social-choice perspective on constitutional-style alignment, we treat principles as voters that evaluate pairs of model responses by judging which better adheres to the relevant principle. 
From this perspective, a minimal requirement - ``axiom'' - for any aggregation method is Pareto Optimality (PO) (also known as Unanimity \citep{arrow1951social}): e.g., in the HHH framework, if response A is more helpful, more honest, and more harmless than B, then A should receive higher reward and thus be more likely to be generated by the model. 
We study a setting in which principles (such as HHH) are modeled as linear directions in a representation space. This is a simplifying assumption motivated by the Linear Representation Hypothesis (LRH) \citep{park2023linear}, which posits that some high-level concepts are well-approximated by linear features in learned embeddings and which has been operationally useful in mechanistic interpretability.

Formally, most of our results lie within the \textit{linear social-choice} framework of \citet{ge2024axioms}: utilities over a fixed representation are linear, pairwise comparisons yield binary labels, and optimizing those labels learns a linear reward model. Although the linear social-choice framework was introduced as a model for RLHF, we contend that both the linearity assumption and axioms such as PO are more realistic and salient in constitutional-style alignment with a small number of principles.\footnote{With many raters, agreement on a binary comparison becomes increasingly unlikely.}

Within the linear social choice model, \citet{ge2024axioms} show that, perhaps surprisingly, an optimally trained linear reward can violate Pareto optimality (PO): everyone prefers response A to response B, yet B receives higher reward. The concrete counterexample involves only two (weighted) principles, thus covering the HHH scenario; we discuss it as a warm-up in \Cref{sec:revisit}. Finding this result counterintuitive, we ask:

\begin{center}
\itshape
How robust are axiomatic violations under loss-based rules?
\end{center}
\subsection{Our Contribution}
We revisit the axiomatic violations through a lens which is compatible with training pipelines. As part of our warm-up in \Cref{sec:revisit}, we provide new intuition on why the PO violation occurs in the first place, by providing a simplified minimal example that clarifies what goes wrong. 
While \citet{ge2024axioms} show that a social choice combinatorial approach can recover axiomatic guarantees, our paper focuses on approaches to obtain guarantees that align with modern ML pipelines—namely, loss-based rules. Our main contributions are three-fold.

\begin{enumerate}

\item \textbf{Beyond linear rewards.} A natural hypothesis is that linear rewards with a frozen embedding are simply too restrictive; perhaps keeping voters linear but using a more expressive reward model, such as bounded-degree polynomials or multilayer perceptrons (MLPs), restores PO. We show that PO and Pairwise Majority Consistency (PMC) can still fail for polynomial rewards of any fixed degree (\Cref{thm:poly-PO}). We further extend the PO impossibility to ReLU MLPs with one hidden layer: for every fixed width \(d\geq2\), there is a finite parameter bound \(W\) for which PO can still fail (\Cref{thm:MLP-PO}).

\item \textbf{PO beyond observed comparisons.} Reward models evaluate responses beyond their training data, so axiomatic guarantees should also apply to unseen comparisons. In \Cref{sec:sampling}, we extend PO from a fixed candidate set to the entire embedding space. Since linear voters' preferences depend only on the direction of the difference between candidate embeddings, we formulate PO over all comparison directions, including those absent from the training data.

\item \textbf{Data-centric perspective.} While social-choice axiomatic analyses typically seek guarantees across all instances, reward-model training also allows us to influence the comparison data through curation and sampling. We therefore ask whether controlling the distribution of comparisons can restore PO while retaining loss-based training. For linear rewards, we establish PO under suitable assumptions in an idealized setting with uniformly distributed comparison directions and equal comparison lengths. We also show that PMC can still fail in this setting.

These results motivate studying data curation and reweighting as ways to recover axiomatic guarantees within existing training pipelines.
\end{enumerate}
Results marked with (\deferredsym) have their proofs deferred to the appendix.
\subsection{Related Work}

The field of social choice \citep{arrow2010handbook, brandt2016handbook} offers a long tradition of axiomatic guarantees 
that provide a lens through which to compare different aggregation rules. This perspective is directly relevant to preference-based alignment: Reinforcement Learning from Human Feedback (RLHF) \citep{christiano2017deep} and Nash Learning from Human Feedback (NLHF) \citep{munos2023nash} have natural analogues in social choice.
The optimal solution to the Bradley–Terry loss in RLHF is known to induce a ranking that is equivalent to the Borda ranking from social choice theory~\citep{Anderson2009BordaZermelo,siththaranjan2023distributional}.
Formally, NLHF coincides with the von Neumann winner defined in~\cite{Dudik2015}, which in turn matches an old rule from social choice, Fishburn's maximal lotteries~\citep{fishburn1984probabilistic}.\footnote{Both concepts are Nash equilibria of games and coincide because the corresponding payoff matrices are related by a positive affine transformation; see~\citep{WangLiuJin2023,maurarivero2025} for the explicit reference to maximal lotteries.}
Building on this connection, several position papers have recently argued for social choice as a useful lens on RLHF and alignment more broadly \citep{conitzer2024social, dai2024mapping}.\\
Among the first wave of technical results that has emerged since, most relevant to us is the work by \citet{ge2024axioms} who propose the linear social-choice model where voters are linear directions over a fixed embedding. Building on prior work by \citep{noothigattu2020axioms} which explores the axiomatic properties of reward functions defined as MLE estimators of underlying random utility models, they analyze the ranking over a fixed candidate set induced by an optimal linear reward with respect to axioms including Pareto optimality (PO) and Pairwise Majority Consistency (PMC). Because these axioms can still be violated in that setting, they adapt a combinatorial social-choice rule to the linear social-choice model to obtain the guarantees. The closest analogue in social choice to linear preferences over a fixed embedding arises in the literature on restricted preference domains (over rankings) \citep{elkind2022preference}. 

\citet{procaccia2025clone} analyze clone robustness and show that an appropriate reweighting of the Bradley--Terry loss can be made to satisfy their axiom. Most existing work at the intersection of alignment and social choice adopts the classic social-choice setting with unstructured alternatives, in contrast to the metric setting we study here. Recent work in this vein includes representative social choice proposed by \citet{qiu2024representative}, the extension of the distortion framework of \citet{procaccia2006distortion} to preference distributions by \citet{golz2025distortion}, an analysis of RLHF when each comparison is labeled by a single annotator by \citet{xiao2025theoretical}, and a proposal towards proportional alignment by \citet{kim2025population}.%

\section{Preliminaries}\label{sec:prelims}

\paragraph{\bf Social choice model.} The set of alternatives is the $d$-dimensional space $\mathbb{R}^d$. A reward function $r: \mathbb{R}^d \to \mathbb{R}$ induces a (weak) ordering $\preccurlyeq_r$ over the set of all alternatives $\mathbb{R}^d$ by ranking alternatives according to their rewards, namely $a \preccurlyeq_r b \iff r(a) \leq r(b)$ for all $a,b \in \mathbb{R}^d$.

There are $n$ voters, and each voter $i$ has a (weak) ordering $\preccurlyeq_i$ over the set of alternatives $\mathbb{R}^d$. Unlike in the usual social choice setting, we do not observe the full orderings, but instead we are only offered a partial view of each voter's preferences. For each voter $i$, we observe a list of pairwise (strict) comparisons. Formally, for each $i$ we are given a nonempty finite set $P_i \subset (\mathbb{R}^d)^2$ that satisfies $(a,b) \in P_i \implies a \prec_i b$. A natural special case is when there is a set $C$ of $m$ candidates in $\mathbb{R}^d$ and the sets $P_i$ contain all\footnote{We assume that the set of candidates $C$ is such that no two candidates are tied for any of the $n$ voters, i.e., for any $a, b \in C$ either $a \prec_i b$ or $b \prec_i a$ holds.} pairwise comparisons between candidates in $C$, i.e., $P_i = \{(a,b) \in C^2: a \prec_i b\}$.

In this context, a voting rule takes as input sets of pairwise comparisons $P_1, \dots, P_n$, one for each voter, and outputs a reward function whose induced ordering attempts to aggregate the voter preferences.

\paragraph{\bf Loss-based voting rules.}
We restrict our attention to voting rules that output a reward function minimizing a total loss function. Given a particular loss function $\ell: \mathbb{R} \to \mathbb{R}$, we define the total loss incurred by reward function $r$ as
$$\mathcal{L}(r) := \sum_{i \in [n]} \sum_{(a,b) \in P_i} \ell(r(a) - r(b))$$
where $P_1, \dots, P_n$ are the sets of pairwise comparisons of the $n$ voters, as described above. The voting rule defined by loss function $\ell$ outputs a reward function $r$ minimizing the total loss $\mathcal{L}$. 

We obtain the Bradley-Terry loss \citep{bradley1952rank, zermelo1929turnier} $\mathcal{L}_{BT}$ used in RLHF by choosing the loss function $\ell$ to be the cross-entropy loss, i.e., $\ell_{\mathrm{BT}}(x) :=\log\bigl(1 + e^{x}\bigr)$.

We will mention the precise assumptions on the loss function $\ell$ in the theorem statements later, but for now let us just think of $\ell$ as a strictly increasing function. Then the idea is that if some voter has ranked $a$ over $b$ in their set $P_i$, a reward function should be penalized for ranking $b$ over $a$ (i.e., for assigning a higher reward to $b$ than $a$).

\paragraph{\bf Linear social choice.}
We will mostly focus on the linear social choice model, where the reward functions are linear, i.e, of the form $r_\theta(x) = \inner{\theta}{x}$. In particular, in this context, a loss-based voting rule will minimize the total loss $\mathcal{L}$ over the set of all linear reward functions. Furthermore, the voters are also assumed to have orderings that are induced by linear reward functions.

\paragraph{\bf Axioms.}
In this paper we will focus on the following two natural axioms.

\begin{definition}[Pareto Optimality (PO)]
A loss-based voting rule satisfies Pareto optimality (PO), if it outputs a reward function $r$ that ranks any comparison $(a,b) \in \cap_i P_i$ correctly, i.e., $r(a) < r(b)$.
\end{definition}

In other words, if there exists a comparison between two candidates on which all voters agree, then PO requires the voting rule to also rank these candidates in the same way as the voters.

\begin{definition}[Pairwise Majority Consistency (PMC)]
An ordering $\prec$ over all alternatives appearing in the instance is a PMC ordering, if it satisfies: $a \prec b$ if and only if a strict majority of voters rank $a$ below $b$, i.e., $|\{i: (a,b) \in P_i\}| > n/2$. A loss-based voting rule satisfies PMC, if whenever a PMC ordering $\prec$ exists, it outputs a reward function that induces $\prec$.
\end{definition}

Note that if a PMC ordering exists, then it is necessarily unique.

\section{Revisiting the Pareto Optimality Violation}\label{sec:revisit}

The counterexample to PO provided by \cite{ge2024axioms} is constructed in \(\mathbb{R}^2\). It uses two (weighted) voters with direction vectors \(v_1 = (1,1)\) and \(v_2 = (-1,0)\) (their magnitudes are irrelevant) and six candidates in \(\mathbb{R}^2\), arranged as two triples. The key idea is to place one triple ($a$ at $(2,1)$, $b$ at $(1,1)$ and $c$ at $(0,0)$) so that the two voters disagree on the induced ordering of its three candidates; with suitable weights, however, the optimal direction is \((1,0)\).

Each candidate is then duplicated and the copies are perturbed within an \(\varepsilon\)-neighborhood of the originals. This replicates every pairwise comparison fourfold and adds an almost constant term to the loss from comparing each candidate with its own copy. A continuity argument shows that for sufficiently small \(\varepsilon\) the optimal linear reward for the perturbed instance remains close to the original one, since as \(\varepsilon \to 0\) the new loss is a positive affine transformation of the old loss.
Finally, choose the perturbation of \(c'\) in the left half-plane so that both voters strictly prefer \(c'\) to \(c\), yielding the desired PO violation.

This construction can be tightened to use only four candidates by duplicating a single point, as we do in the proof of \Cref{thm:poly-PO}. For three non-collinear candidates, every assignment of reward differences is realizable by a linear reward. Thus, for the Bradley–Terry objective with complete comparison data, any finite minimizer satisfies PO.
Disagreement between voters (e.g., \(v_1\) prefers \(a \succ b \succ c\) while \(v_2\) prefers the reverse) means that the loss contains both
\[
\log\!\bigl(1+e^{-\inner{\theta}{a-b}}\bigr)
\quad\text{and}\quad
\log\!\bigl(1+e^{-\inner{\theta}{b-a}}\bigr),
\]
(possibly with different weights). 
These terms penalize large projections along disputed comparison directions. When these directions span the embedding space, their combined effect acts as an implicit \emph{norm constraint}.
The same phenomenon holds more generally for losses bounded below but diverging to \(\infty\) as their argument \(\to \infty\).

We next study a minimal example with one voter and three candidates where the optimal linear reward subject to a norm constraint violates PO. 
This instance provides insight into why PO can fail even without such explicit constraints on instances with multiple voters.
Consider a single voter \(v=(\varepsilon,1)\) with \(1>\varepsilon>0\) and three candidates
\(a=(1,0)\), \(b=(0,0)\), \(c=(-\delta,\delta)\) where \(0<\delta\ll \frac{\varepsilon}{1-\varepsilon}\). Constrain \(\theta\) to unit length,
\(\theta=(\theta_1,\ \pm\sqrt{1-\theta_1^2})\) with \(\theta_1\in[-1,1]\). The pairwise Bradley--Terry terms are
\[
\begin{aligned}
(a,b)&:\ \log\!\bigl(1+e^{-\inner{\theta}{a-b}}\bigr)=\log\!\bigl(1+e^{-\theta_1}\bigr),\\[2pt]
(a,c)&:\ \log\!\bigl(1+e^{-\inner{\theta}{a-c}}\bigr)
=\log\!\bigl(1+e^{-\bigl((1+\delta)\theta_1\ \mp\ \delta\sqrt{1-\theta_1^2}\bigr)}\bigr),\\[2pt]
(b,c)&:\ \log\!\bigl(1+e^{-\inner{\theta}{c-b}}\bigr)
=\log\!\bigl(1+e^{\delta(\theta_1\ \mp\ \sqrt{1-\theta_1^2})}\bigr)
\ \le\ \log\!\bigl(1+e^{\delta\sqrt{2}}\bigr),
\end{aligned}
\]
since \(\max_{\theta_1\in[-1,1]}\bigl(\theta_1\pm\sqrt{1-\theta_1^2}\bigr)\leq \sqrt{2}\). Thus the \((b,c)\) term is \(\log2+O(\delta)\), and the
\((a,c)\) term differs from the \((a,b)\) term by \(O(\delta)\),
uniformly over unit vectors \(\theta\). The total loss therefore
converges uniformly, as \(\delta\to0\), to
\[
2\log(1+e^{-\theta_1})+\log2,
\]
whose unique minimizer on the unit circle is \((1,0)\).
By compactness, every minimizer of the total loss approaches
\((1,0)\). Consequently, for sufficiently small \(\delta\),
every minimizer has \(\theta_1>\theta_2\), yielding
\(\inner{\theta}{b}>\inner{\theta}{c}\) while
\(\inner{v}{b}<\inner{v}{c}\).
The example shows that a norm constraint can cause PO violations even with one voter. With multiple voters, disagreement can create a similar restriction implicitly.
Intuitively, the norm constraint can be interpreted as a finite “budget.” Comparisons in more common directions, or with longer difference vectors such as \(a-b\), can contribute more to the loss when ranked incorrectly, giving them greater influence over how this budget is used.

\section{Beyond Linear Reward Functions}\label{sec:genrewards}

We now extend our analysis beyond linear reward functions
and investigate whether more expressive reward classes
can overcome the limitations of loss-based voting rules. We establish two impossibility results. First, in \Cref{subsec:polyrewards}we show
that restricting reward functions to polynomials of any
fixed degree does not guarantee Pareto optimality (PO)
or PMC. We then extend this to a similar impossibility
result in \Cref{subsec:MLPrewards} for one-hidden-layer ReLU networks of any fixed
width.

\subsection{Polynomial Reward Functions}\label{subsec:polyrewards}
We begin by showing that loss-based voting rules can violate both PO and PMC even when the reward functions are polynomials of arbitrarily high but bounded degree.

\begin{restatable}{theorem}{polyPO}\label{thm:poly-PO}
Any loss-based voting rule with a loss function $\ell$ that is strictly convex, lower bounded, and differentiable with $\ell'(0) > 0$, fails to satisfy PO and PMC even with polynomial reward functions of bounded degree. Furthermore, this already holds in two dimensions and with three voters that all lie in the positive quadrant.
\end{restatable}

In particular, this theorem applies to the Bradley--Terry loss, which thus fails to satisfy PO and PMC even with polynomial reward functions of bounded degree.

\subsubsection{Proof of \Cref{thm:poly-PO}}
The instance consists of $m+1 := d(d+1) + 2$ candidates $c_0, c_1, c_2, \dots, c_m$, whose positions in $\mathbb{R}^2$ will be specified later. Furthermore, there are two weighted voters:
\begin{itemize}
\item Voter $v_1 = (1,0)$ has a fraction $\alpha \in (1/2,1)$ of the votes and ranks the candidates in the order
$$c_1 \prec c_0 \prec c_2 \prec c_3 \prec \dots \prec c_m.$$
\item Voter $v_2 = (0,1)$ has a fraction $1-\alpha$ of the votes and ranks the candidates in the order
$$c_m \prec \dots \prec c_1 \prec c_0.$$
\end{itemize}
Note that the voters disagree on all comparisons, except the comparison between $c_0$ and $c_1$, where they agree. The proof works for any $\alpha \in (1/2,1)$, but we can set, e.g., $\alpha = 2/3$ to obtain a setting with three (unweighted) voters.

\paragraph{\bf Total loss function.}
For now, consider allowing arbitrary reward values $r_0, r_1, \dots, r_m \in \mathbb{R}$ for the candidates $c_0, c_1, \dots, c_m$. The total loss of the reward vector $r$ on this instance can be written as
\begin{equation}\label{eq:poly-total-loss}
\begin{split}
\mathcal{L}(r) = \mathcal{L}(r_0,r_1,\dots,r_m) := \alpha &\left( \ell(r_1-r_0) +  \sum_{j=2}^m \ell(r_0-r_j) + \sum_{1 \leq i < j \leq m} \ell(r_i-r_j) \right)\\
&+ (1-\alpha) \sum_{0 \leq i < j \leq m} \ell(r_j-r_i)
\end{split}
\end{equation}
Note that we can assume without loss of generality that, say, $r_1 = 0$, since the total loss does not change if we subtract $r_1$ from all rewards.

\paragraph{\bf Optimal rewards in the degenerate instance.}
In our instance, we will position candidate $c_0$ very close to $c_1$. Thus, the instance will be closely related to a ``degenerate'' instance, where $c_0$ and $c_1$ lie at the same position. In that degenerate instance, the optimal (arbitrary) rewards are given by the following optimization problem:

\begin{equation}\label{eq:poly-opt-rewards}
\begin{split}
\min_r \quad &\mathcal{L}(r)\quad
\textup{s.t.} \quad r_0=r_1 = 0
\end{split}
\end{equation}

\begin{restatable}[\deferredsym]{claim}{uniquerewards}\label{clm:poly-unique-rewards}
   The optimization problem \eqref{eq:poly-opt-rewards} has a unique solution $0 = r_1^* < r_2^* < \dots < r_m^*$. 
\end{restatable}

\paragraph{\bf Positioning of the candidates.}
We use these optimal rewards $0 = r_1^* < r_2^* < \dots < r_m^*$ to define the positions of the candidates. First, we let $c_1 = (0,0)$ and $c_0 = (\delta, \delta)$ for some sufficiently small $\delta > 0$ to be specified later. Next, for each $j \in [d]$, we let $L_j := \{(x,y) \in \mathbb{R}^2: y = -2x + j\}$, i.e., $L_j$ is the line going through point $(0,j)$ with slope $-2$. Then, for each $i = 2, 3, \dots, n$, we position candidate $c_i$ at the unique point $(x,y)$ on line $L_j$ such that $-x-y = r_i^*$, where $j = \lceil (i-1)/(d+1) \rceil$. As a result, for any $i \geq 2$, candidate $c_i$ is positioned at $(x,y)=(r_i^*+j,-2 r_i^*-j)$, where $j = \lceil (i-1)/(d+1) \rceil$.

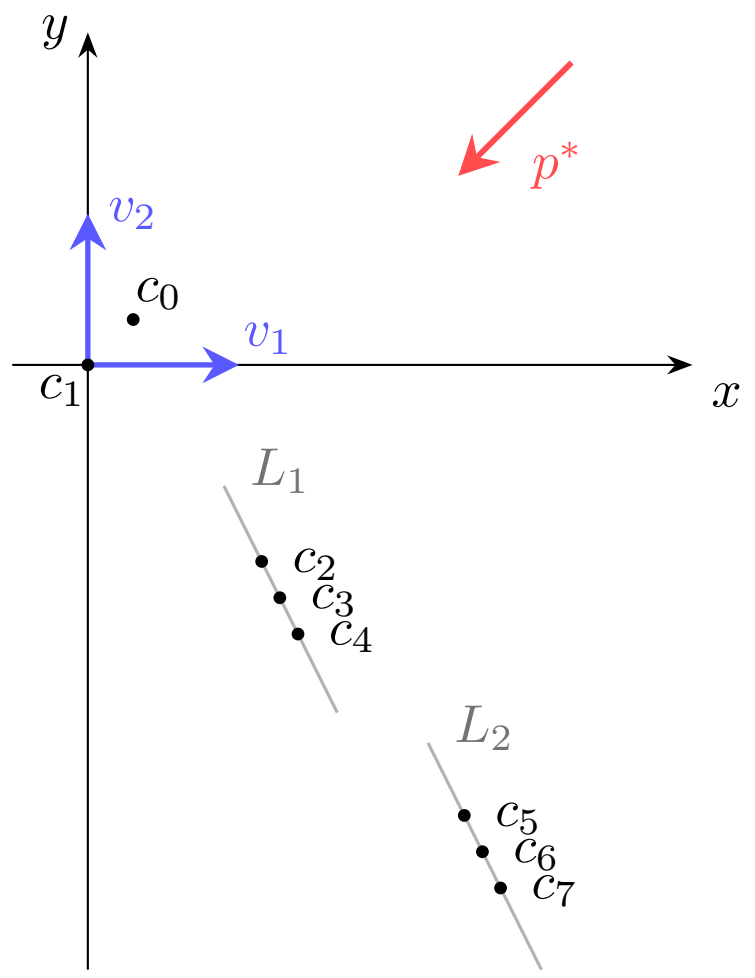
\begin{figure}[t]
  \centering
  \begin{tikzpicture}[scale=1.0, >=Stealth]
\small

\def\dval{2}
\def\delta{0.30}
\def\xmin{-0.5}\def\xmax{4.0}
\def\ymin{-4.0}\def\ymax{2.2}

\def\LoneYmin{-2.3} \def\LoneYmax{-0.8}
\def\LtwoYmin{-4} \def\LtwoYmax{-2.5}

\tikzset{
  axis/.style   = {thin, black},
  line/.style  = {line width=0.6pt, black!30},
  voter/.style = {line width=1.0pt, blue!65, -{Stealth[length=2.4mm,width=2.4mm]}},
  pdir/.style   = {line width=1.1pt, red!70, -{Stealth[length=2.6mm,width=2.6mm]}},
  cand/.style= {draw=black, fill=black, circle, inner sep=1.0pt}
}

\newcommand{\placecandidate}[3]{%
  \pgfmathsetmacro{\xx}{#1 + #2}
  \pgfmathsetmacro{\yy}{-#1 - 2*(#2)}
  \fill[cand] (\xx,\yy) circle (1.0pt);
  \node[anchor=west] at (\xx+0.08,\yy-0.02) {$#3$};
}

\draw[axis,->] (\xmin,0)--(\xmax,0) node[below right] {$x$};
\draw[axis,->] (0,\ymin)--(0,\ymax) node[left] {$y$};

\pgfmathsetmacro{\xAone}{(1 - \LoneYmin)/2}
\pgfmathsetmacro{\xBone}{(1 - \LoneYmax)/2}
\draw[line] (\xAone,\LoneYmin) -- (\xBone,\LoneYmax);
\node[black!55, anchor=west] at (0.95,-0.7) {$L_{1}$};

\pgfmathsetmacro{\xAtwo}{(2 - \LtwoYmin)/2}
\pgfmathsetmacro{\xBtwo}{(2 - \LtwoYmax)/2}
\draw[line] (\xAtwo,\LtwoYmin) -- (\xBtwo,\LtwoYmax);
\node[black!55, anchor=west] at (2.3,-2.4) {$L_{2}$};

\draw[voter] (0,0) -- (1,0) node[anchor=west, text=blue!65, xshift=-3pt, yshift=5pt] {$v_1$};
\draw[voter] (0,0) -- (0,1) node[anchor=west, text=blue!65] {$v_2$};

\fill[cand] (0,0) circle (1.0pt) node[anchor=north east] {$c_1$};
\fill[cand] (\delta,\delta) circle (1.0pt) node[anchor=south west] {$c_0$};

\draw[pdir] (3.2, 2.0) -- ++(-0.75, -0.75)
  node[anchor=west, xshift=10pt, yshift=2pt] {$p^*$};

\placecandidate{1}{0.15}{c_2}
\placecandidate{1}{0.27}{c_3}
\placecandidate{1}{0.39}{c_4}
\placecandidate{2}{0.49}{c_5}
\placecandidate{2}{0.61}{c_6}
\placecandidate{2}{0.73}{c_7}

\end{tikzpicture}
  \caption{Illustration of the positioning of the candidates for degree $d=2$. The red arrow labeled $p^*$ indicates the direction of increase of the linear polynomial $p^*(x,y)=-x-y$.}
  \label{fig:poly-positioning}
\end{figure}

We give a more detailed description of the positioning of candidates $c_i$ for $i \geq 2$. We let $c_2$ be the unique point $(x,y)$ on line $L_1$ such that $-x-y = r_2^*$. Similarly, we let $c_3$ be the unique point $(x,y)$ on line $L_1$ such that $-x-y = r_3^*$. We continue positioning points on $L_1$ in this manner, until $d+1$ points have been positioned. Then, we switch to $L_2$. In other words, $c_{d+3}$ is the unique point $(x,y)$ on line $L_2$ such that $-x-y = r_{d+3}^*$. We proceed in this manner, placing $d+1$ points on each line, and then moving on to the next line. Note that the number of candidates to be positioned (without $c_0$ and $c_1$, since these have been fixed above) is exactly $m-1 = d(d+1)$, so we are able to position exactly $d+1$ candidates on each of the $d$ lines $L_1, \dots, L_d$. See \Cref{fig:poly-positioning} for an illustration. By construction the following holds.

\begin{restatable}{claim}{polypos}\label{clm:poly-positioning}
The positions of the candidates $c_0, \dots, c_m$ are consistent with the rankings of voters $v_1$ and $v_2$. Furthermore, the polynomial $p^*(x,y) = -x-y$ satisfies $p^*(c_i) = r_i^*$ for all $i \in [m]$ and ranks the candidates in the order $c_0 \prec c_1 \prec c_2 \prec \dots \prec c_m$.
\end{restatable}

\paragraph{\bf Degree-$d$ polynomial reward functions.} We consider the class of reward functions that are polynomials of degree at most $d$. As before, without loss of generality, we can restrict our attention to polynomials $p$ that satisfy $p(c_1) = 0$, i.e., the constant term is zero (since $c_1$ lies at the origin). The optimal such polynomial reward functions are given by the following optimization problem:
\begin{equation}\label{eq:poly-opt}
\begin{split}
\min_p \quad &\mathcal{L}(p(c_0),p(c_1),\dots,p(c_m)) \\
\textup{s.t.} \quad &\text{$p(x,y)$ polynomial of degree at most $d$} \\
&p(0,0) = 0
\end{split}
\end{equation}

As shown in \Cref{clm:poly-positioning}, the polynomial $p^*(x,y) = -x-y$ ranks $c_1$ above $c_0$ when used as a reward function, i.e., $p^*(c_1) > p^*(c_0)$. On the other hand, both voters $v_1$ and $v_2$ rank $c_0$ over $c_1$. In the rest of this proof, our goal will be to show that the optimal polynomial degree-$d$ reward function for our instance (i.e., any solution to \eqref{eq:poly-opt}) is close to $p^*$, and thus also ranks $c_1$ above $c_0$. This will immediately show that Pareto optimality does not hold for our instance. %

\begin{claim}\label{clm:poly-unique}
In the degenerate instance where $\delta = 0$, and thus $c_0 = c_1 = (0,0)$, the polynomial $p^*(x,y) = -x-y$ is the unique optimal solution of \eqref{eq:poly-opt}.
\end{claim}

\begin{proof}
By \Cref{clm:poly-positioning}, the polynomial $p^*$ achieves the optimal rewards $r^*$ from \eqref{eq:poly-opt-rewards}, and thus $p^*$ is an optimal solution of \eqref{eq:poly-opt} in the degenerate instance where $\delta = 0$. In order to show that $p^*$ is the unique optimal solution, it suffices to show that no other degree-$d$ polynomial achieves the optimal rewards $r^*$, which are the unique optimal solution of \eqref{eq:poly-opt-rewards} by \Cref{clm:poly-unique-rewards}. This follows from the fact that the zero polynomial is the only degree-$d$ polynomial that simultaneously vanishes at all points~$c_i$, a fact which we prove next.

Consider a polynomial $p$ of degree at most $d$ such that $p(c_i) = 0$ for all $i \in [m]$. We will show that $p=0$. First, we apply a rotation around the origin to the $(x,y)$ plane such that the lines $L_j$ are now of the form $L_j = \{(x,y) \in \mathbb{R}^2: y = s_j\}$ for some $0 < s_1 < s_2 < \dots < s_d$. Note that $p$ is still a polynomial of degree at most $d$ in this new basis.

Now since $p$ vanishes at $d+1$ distinct points on $L_j$, it follows that $p$ restricted to $y=s_j$ is the zero polynomial, i.e., $p\big|_{y=s_j} = 0$. As a result,\footnote{More formally, divide polynomial $p$ by polynomial $y-s_1$ in $(\mathbb{R}[x])[y]$. We obtain $p(x,y) = (y-s_1)q(x,y) + r(x)$, where we note that $r$ only depends on $x$ since it must have degree strictly less than one in $y$. Now $p\big|_{y=s_1} = 0$ implies that $r=0$. Then, we can continue by dividing $q$ by $y-s_2$ in the same manner to obtain the full factorization.} $p$ must contain a factor $(y-s_j)$ for each $j \in [d]$. Since $p$ has degree at most $d$, it follows that $p$ can be written as $p(x,y) = C \cdot \prod_{j \in [d]} (y - s_j)$. But now $p(0,0) = r_1^* = 0$, together with $s_j \neq 0$ for all $j$, implies that $C=0$ and thus $p=0$.
\end{proof}

We will use Berge's maximum theorem to argue that for sufficiently small $\delta > 0$, any optimal solution to \eqref{eq:poly-opt} must be close to $p^*$.

\begin{theorem}[Berge's Maximum Theorem \citep{berge1997topological}; simplified version]\label{thm:berge}
Let $A \subseteq \mathbb{R}^n$ and $B \subseteq \mathbb{R}^m$ such that $B$ is nonempty and compact. Let $f \colon A \times B \rightarrow \mathbb{R}$ be continuous. Define the set-valued function $f^* \colon A \rightrightarrows B$ by $f^*(a) = \argmax_{b \in B} f(a,b)$. Then $f^*$ is upper-hemicontinuous with nonempty and compact values.
\end{theorem}

\begin{claim}\label{clm:poly-limit}
There exists a sufficiently small $\delta > 0$ such that any optimal solution of \eqref{eq:poly-opt} satisfies $p(c_1) > p(c_0)$.
\end{claim}

Before proving this claim, let us see why this implies \Cref{thm:poly-PO}. For $\delta$ as given by the claim, any optimal solution ranks $c_1$ over $c_0$. On the other hand, both voters $v_1$ and $v_2$ rank $c_0$ over $c_1$, so Pareto optimality is not satisfied.
Furthermore, voter $v_1$ always (trivially) agrees with a strict majority of the voters; in particular, the ordering of voter $v_1$ is PMC for the instance. However, any optimal solution ranks $c_1$ over $c_0$, even though a majority (in fact, everyone) agrees to the opposite. Thus, the example also fails PMC.

\begin{proof}
In order to use Berge's maximum theorem, we need to make sure that the domain over which we are optimizing is compact. Let $S = \{s = (s_{i,j})_{i,j \geq 0: i+j \leq d}: s_{i,j} \in \mathbb{R}, s_{0,0} = 0\}$ represent all polynomials of degree at most $d$ with zero constant coefficient, i.e., for each $s \in S$, the corresponding polynomial is given by $p_s(x,y) = \sum_{i=0}^d\sum_{j=0}^{d-i} s_{i,j }x^iy^j$. Let $S'$ denote the subset of $S$ where all coefficients have magnitude at most $2$, i.e., $S' = \{s \in S: |s_{i,j}| \leq 2\}$. Let $s^* \in S$ be such that $p_{s^*} = p^*$, i.e., $s^*_{1,0} = s^*_{0,1} = -1$ and all other coefficients are zero. Note that $S'$ is compact and $s^*$ lies in the interior of $S'$.

Now define $f: [0,1] \times S' \to \mathbb{R}$, $(\delta, s) \mapsto \mathcal{L}(p_s(c_0(\delta)),p_s(c_1),\dots,p_s(c_m))$. Clearly, $f$ is continuous, since the loss function $\ell$ is continuous, and $S'$ is nonempty and compact. Thus, by Berge's maximum\footnote{The theorem clearly also applies to minimization by replacing $f$ by $-f$.} theorem (\Cref{thm:berge}), the set-valued function $f^* \colon [0,1] \rightrightarrows S', \delta \mapsto \argmin_{s \in S'} f(\delta,s)$ is upper-hemicontinuous.

Recall that by \Cref{clm:poly-unique}, when $\delta = 0$, the polynomial $p^*(x,y) = -x-y$ is the unique minimizer of $\mathcal{L}$, i.e., $f^*(0) = \{s^*\}$, where $s^* \in S$ is as defined above such that $p^* = p_{s^*}$. Since $f^*$ is upper-hemicontinuous, there exists $\eps > 0$ such that for all $\delta \in [0,\eps]$, we have $\|s-s^*\|_\infty \leq 1/2$ for all $s \in f^*(\delta)$. In particular, this means that for all $\delta \in [0,\eps]$, $f^*(\delta)$ lies in the interior of $S'$. By convexity of the function $s \mapsto \mathcal{L}(p_s(c_0(\delta)),p_s(c_1),\dots,p_s(c_m))$, it follows that $f^*(\delta)$ is the set of minimizers over all of $S$ (not just $S'$). Thus, whenever $\delta \in [0,\eps]$, any minimizer $p_s$ of \eqref{eq:poly-opt} satisfies $\|s-s^*\|_\infty \leq 1/2$. But this implies that $p_s(c_1) = p_s(0,0) = 0$ and $p_s(c_0) = p_s(\delta,\delta) \leq - \delta/2 - \delta/2 + O(\delta^2) = -\delta + O(\delta^2)$. As a result, there exists a sufficiently small $\delta > 0$ such that $p_s(c_1) > p_s(c_0)$. So for sufficiently small $\delta > 0$, any minimizer $p$ of \eqref{eq:poly-opt} must satisfy $p(c_1) > p(c_0)$.
\end{proof}

\subsection{MLP Reward Functions}\label{subsec:MLPrewards}

In this section, we go a step further and study reward models parameterized by simple multilayer perceptrons (MLPs). While similar to polynomial rewards in some respects, MLPs are universal approximators and arguably more representative of the reward models used in alignment practice. We consider a one-hidden-layer ReLU MLP of width at most $d$ with input in
$\mathbb R^2$ which is a function $f:\mathbb R^2\to\mathbb R$ of the form
\begin{equation}\label{MLP}
f(x,y)
=
\beta+\sum_{i=1}^d\alpha_i[\psi_i(x,y)]_+,
\qquad
\psi_i(x,y):=a_i x+b_i y+\eta_i,
\end{equation}
where $[t]_+:=\max\{0,t\}$. Networks of width smaller than $d$ are
included by allowing some $\alpha_i$ to be zero.

\begin{theorem}\label{thm:MLP-PO}
Let $\ell$ be strictly convex, lower bounded, and differentiable with $\ell'(0)>0$. For every $d\ge2$, there exists $W<\infty$ such that any loss-based voting rule using $\ell$ fails PO  even if reward functions are one-hidden-layer width-$d$ ReLU networks with parameters in $[-W,W]$.

\end{theorem}
Furthermore, similar to \Cref{thm:poly-PO}, \Cref{thm:MLP-PO} already holds in two dimensions and with three voters.

\subsubsection{Proof of \Cref{thm:MLP-PO}}
We first construct a degenerate instance and a width-$d$ network that realizes its optimal rewards. The candidate placement forces every optimal network to agree locally near the pair we subsequently perturb. We then show that, after a sufficiently small perturbation, all voters prefer one candidate to the other, while every optimal network ranks them in the opposite order.

The instance has $m+1=O(d^3)$ candidates $c_0,c_1,\dots,c_m$ in $\mathbb R^2$, whose positions we specify below, and two weighted voters:
\begin{itemize}
\item Voter $v_1 = (1,0)$ has a fraction $\alpha \in (1/2,1)$ of the votes and ranks the candidates in the order
$$c_0 \prec_1 c_2 \prec_1 c_1 \prec_1 c_3 \prec_1 \dots \prec_1 c_m.$$
\item Voter $v_2 = (0,1)$ has a fraction $1-\alpha$ of the votes and ranks the candidates in the order
$$c_m \prec_2 \dots c_2 \prec_2 c_1 \prec_2 c_0.$$
\end{itemize}
Note that the voters disagree on all comparisons, except the comparison between $c_1$ and $c_2$, where they agree. The proof works for any $\alpha \in (1/2,1)$, but we can set, e.g., $\alpha = 2/3$ to obtain a setting with three (unweighted) voters.

\paragraph{\bf Total loss function.}
For now, consider allowing arbitrary reward values $r_0, r_1, \dots, r_m \in \mathbb{R}$ for the candidates $c_0, c_1, \dots, c_m$. The total loss of the reward vector $r$ on this instance can be written as
\begin{equation}\label{eq:MLP-total-loss}
\begin{split}
\mathcal{L}(r) = \mathcal{L}(r_0,r_1,\dots,r_m) := \alpha &\left( \ell(r_2-r_1) +
\sum_{\substack{0 \le i < j \le m \\ i \ne 1 \ \text{or}\ j \ne 2}} \ell(r_i-r_j)
 \right)\\
&+ (1-\alpha) \sum_{0 \leq i < j \leq m} \ell(r_j-r_i)
\end{split}
\end{equation}
Note that we can assume without loss of generality that, say, $r_0 = 0$ (or any other constant), since the total loss does not change if we subtract $r_0$ from all rewards.

\paragraph{\bf Optimal rewards in the degenerate instance.}
In our instance, we will position candidate $c_1$ very close to $c_2$. Thus, the instance will be closely related to a ``degenerate'' instance, where $c_1$ and $c_2$ lie at the same position. In that degenerate instance, the optimal (arbitrary) rewards are given by solutions to the following optimization problem:
\begin{equation}\label{eq:MLP-opt-rewards}
\begin{split}
\min_r \quad &\mathcal{L}(r) \\
\textup{s.t.} \quad &r_1=r_2
\end{split}
\end{equation}

\begin{restatable}{claim}{MLPuniqueRewards}
\label{clm:MLP-unique-rewards}
After normalizing $r_0^*=0$, the degenerate optimization problem has a
unique solution satisfying
$$
0=r_0^*<r_1^*=r_2^*<r_3^*<\cdots<r_m^*.
$$
\end{restatable}

\paragraph{\bf Positioning of the candidates.}

Define the linear map
$\varphi:\mathbb R^2\to\mathbb R$ by
$\varphi(c):=\langle\one,c\rangle.$
Thus, for $c=(x,y)$, we have $\varphi(c)=x+y$.
We first describe the construction in one dimension. Normalize the optimal degenerate reward vector so that $r_0^*=0<r_1^*=r_2^*<r_3^*<\cdots<r_m^*$.

Let $q:=1+d+\binom d2$, let $k:=3q$, and let $m:=dk+2$. We partition the indices $3,\dots,m$ into $d$ consecutive blocks $I_1,\dots,I_d$, where $I_i:=\{(i-1)k+3,\dots,ik+2\}$. Thus each block has size $k$. For each $i\in[d]$, let $C_i:=\{c_j:j\in I_i\}$ denote the corresponding set of candidates. The purpose of the block $I_i$ is that all candidates with indices in $I_i$ will be placed on the $i$-th nonconstant affine piece of the constructed spline. The large value of $k$ is used later to force, for any competing width-$d$ MLP, three candidates from the same block to lie in a common region of linearity.

We now construct a one-dimensional continuous piecewise-affine function $g:\mathbb R\to\mathbb R$. Set $t_1:=0$ and $s_1:=0$. For $i=2,\dots,d$, define $$t_i:=\frac{1}{2}(r^*_{(i-1)k+2}+r^*_{(i-1)k+3}).$$ Thus $t_i$ lies strictly between the last reward value in block $I_{i-1}$ and the first reward value in block $I_i$. We define the breakpoint locations recursively by $$s_i:=s_{i-1}-\frac{t_i-t_{i-1}}{M^{i-2}}$$ for $i=2,\dots,d$, where $M>0$ will be chosen later in the proof. Since $t_i>t_{i-1}$, we have $s_1>s_2>\cdots>s_d$. The interval between $s_i$ and $s_{i+1}$ is chosen so that a line of slope $-M^{i-1}$ increases from value $t_i$ to value $t_{i+1}$.

Now define $g:\mathbb R\to\mathbb R$ by
$$
g(s)
:=
\begin{cases}
t_1, & s\ge s_1,\\[2pt]
t_i-M^{i-1}(s-s_i), & s\in [s_{i+1},s_i]\quad \text{for } i=1,\dots,d-1,\\[2pt]
t_d-M^{d-1}(s-s_d), & s\le s_d.
\end{cases}
$$ Then $g$ is continuous, piecewise affine, satisfies $g(s_i)=t_i$ for all $i=1,\dots,d$, and has slopes $-1,-M,\dots,-M^{d-1}$ on its nonconstant affine pieces.

We then define $z:\mathbb R^2\to\mathbb R$ by
$z(c):=g(\varphi(c)).$ Thus the hinge lines of $z$ are the parallel lines
$H_i:=\{c\in\mathbb R^2:\varphi(c)=s_i\}, i\in[d],$
and on the $i$-th nonconstant strip the gradient of $z$ is
$-M^{i-1}\one$.

For each $i\in[d]$ and each $j\in I_i$, define
$\gamma_j
:=
s_i-\frac{r_j^*-t_i}{M^{i-1}}.$ We define $\gamma_j$ to be the point on the $i$-th affine piece of $g$ where $g$ takes value $r_j^*$. If $i<d$, then $\gamma_j\in (s_{i+1},s_i)$, while if $i=d$, then $\gamma_j<s_d$. Therefore every candidate with index in $I_i$ is placed on the $i$-th nonconstant affine piece of $z$, as desired.
Specifically, we place $c_j$ on the level line
$
\Lambda_j
:=
\{c\in\mathbb R^2:\varphi(c)=\gamma_j\}.
$
Thus
$\varphi(c_j)=\gamma_j$ and $z(c_j)=g(\gamma_j)=r_j^*.$
 We first fix $
c_0=(0,0), c_2=\left(\frac{r_1^*}{2},-\frac{3r_1^*}{2}\right).$
By construction,
$\varphi(c_0)=0$ and $\varphi(c_2)=-r_1^*.$

For $j\geq3$, write $c_j=(x_j,y_j),y_j:=\gamma_j-x_j.$
Then $\varphi(c_j)=x_j+y_j=\gamma_j$, so
$c_j\in\Lambda_j$. The numbers $x_j$ are chosen inductively. Suppose that
$c_0,c_2,c_3,\dots,c_{j-1}$
have already been chosen, with no three of these points collinear. Let
$$
\mathcal L_{j-1}
:=
\{L_{a,b}:a,b\in\{c_0,c_2,c_3,\dots,c_{j-1}\},\ a\neq b\},
$$
where $L_{a,b}$ denotes the line through $a$ and $b$. Define
$F_j
:=
\Lambda_j\cap \bigcup_{L\in\mathcal L_{j-1}}L.$
No two distinct previously chosen candidates lie on the same level line
$x+y=\gamma$, and hence no line in $\mathcal L_{j-1}$ is parallel to
$\Lambda_j$. Therefore each line in $\mathcal L_{j-1}$ intersects
$\Lambda_j$ in at most one point, so $F_j$ is finite.

We choose
$c_j=(x_j,\gamma_j-x_j)\in\Lambda_j\setminus F_j$
with $x_j$ sufficiently large so that $x_j>x_\ell$ and $y_j<y_\ell$
for every previously chosen candidate $c_\ell$. %

By induction, no three candidates in $\{c_0,c_2,c_3,\dots,c_m\}$ are collinear, and the $x$-coordinates of
$c_2,c_3,\dots,c_m$ are strictly increasing while the corresponding $y$-coordinates are strictly
decreasing.
It remains to choose the perturbed copy $c_1$. Let
$c_1(\delta):=c_2+\delta(1,2).$ There are only finitely many values of $\delta$ for which $c_1(\delta)$
lies on a line through two of the already chosen candidates. We choose
$\delta>0$ sufficiently small and outside this finite set. We also require
$\delta<\frac{3r_1^*}{4}$
and
$\delta<x(c_3)-x(c_2),$
so that $c_1$ lies between $c_2$ and $c_3$ in the $x$-coordinate
order and between $c_2$ and $c_0$ in the $y$-coordinate order. Thus
$$
x(c_0)<x(c_2)<x(c_1)<x(c_3)<\cdots<x(c_m),
$$
and
$$
y(c_m)<\cdots<y(c_3)<y(c_2)<y(c_1)<y(c_0).
$$
With this choice, no three points among $c_0,c_1,\dots,c_m$ are
collinear, and their coordinates are consistent with the voter rankings.

Moreover, $\varphi(c_1)=\varphi(c_2)+3\delta=-r_1^*+3\delta$.
For sufficiently small $\delta>0$, we have
$\varphi(c_1)\in[s_2,s_1].$
Hence $c_1$ lies in the first nonconstant affine strip of $z$. Since
$g(s)=-s$ for $s\in[s_2,s_1]$, it follows that
$
z(c_1)
=
g(\varphi(c_1))
=
r_1^*-3\delta
<
r_1^*
=
z(c_2).
$
Thus the constructed reward function $z$ ranks $c_2$ above $c_1$,
while both voters rank $c_1$ above $c_2$.
In the following we will refer to the above instance in the case when $\delta=0$ as the degenerate instance.

\paragraph{\bf Realization by a width-$d$ MLP.}
Recall the form of the ReLU MLPs we consider from \eqref{MLP}.
Let $\Theta_d:=[-W,W]^{1+4d},$ and write each such an MLPs $f_{\theta}$'s parameter vector as
$
\theta
=
\bigl(
\beta,
(\alpha_i,a_i,b_i,\eta_i)_{i=1}^d
\bigr)
\in\Theta_d.$ The optimal MLPs are the solutions of
\begin{equation}\label{eq:MLP-opt}
\min_{\theta\in\Theta_d}
\mathcal L\bigl(
f_\theta(c_0),f_\theta(c_1),\dots,f_\theta(c_m)
\bigr).
\end{equation}
We next observe that the constructed function $z$ can be  written as
width-$d$ MLP for some $\theta$. For $i\in[d]$, define
$\psi_i(x,y)=s_i-(x+y).$
Then
$$
z(x,y)
=
\sum_{i=1}^d \alpha_i[\psi_i(x,y)]_+,
\qquad\text{where }
\alpha_1:=1,
\alpha_i:=M^{i-1}-M^{i-2}
\quad\text{for }i=2,\dots,d.
$$
Thus the corresponding MLP parameters are $a_i=b_i=-1, \eta_i=s_i$ and $ \beta=0.$

Finally, we choose
$$W\geq\max\left\{
1,\ \max_{i\in[d]}|s_i|,\
\max_{2\leq i\leq d}(M^{i-1}-M^{i-2})
\right\}.
$$ With this choice of $W$, the parameter vector of $z$ belongs to $\Theta_d$,
so $z$ is feasible for \eqref{eq:MLP-opt}.
Note that this choice of $W$ is independent of $\delta$. 
In the degenerate instance, $z$ realizes the unrestricted optimal reward vector $r^*$. Let $f^*$ be any optimal solution to the constrained optimization problem~\eqref{eq:MLP-opt}, and define
$\widetilde r_j:=f^*(c_j)-f^*(c_0)$ for $j=0,\ldots,m$.
Then $\widetilde r_0=0$, and since $c_1=c_2$, we also have $\widetilde r_1=\widetilde r_2$. Hence $\widetilde r$ is feasible for the unrestricted optimization problem in the degenerate instance.

Since $z$ is feasible for~\eqref{eq:MLP-opt} and realizes the unrestricted optimal reward vector $r^*$, the constrained and unrestricted optimization problems have the same optimal objective value. Moreover, the loss is invariant under adding a common constant to all rewards, so $\widetilde r$ attains the same objective value as the reward vector induced by $f^*$. Hence $\widetilde r$ is an optimal solution to the unrestricted problem under the normalization $\widetilde r_0=0$. By uniqueness of the unrestricted optimum under this normalization, $\widetilde r_j=r_j^*$ for every $j=0,\ldots,m$. Therefore,
$f^*(c_j)-f^*(c_0)=r_j^*=z(c_j)$ for every $j=0,\ldots,m$.
\paragraph{Properties of any optimal MLP for the degenerate instance.}

Given MLP $f$ as defined in \eqref{MLP}, define the $i$th \textit{hinge line} as
$H_i:=\{(x,y):\psi_i(x,y)=0\}$. Our first Lemma relies on the following useful fact.
\begin{fact}\label{fact:numregions}\citep{zaslavsky1975facing}
The hinge lines $(H_i)_{i\in [d]}$ partition $\mathbb R^2$ into at most
$q_d=1+d+\binom d2$
regions of linearity.
\end{fact}

Throughout the next three lemmas, let $f^*$ be an
arbitrary width-$d$ MLP that is an optimal solution
of \eqref{eq:MLP-opt} for the degenerate instance
$\delta=0$.
\begin{lemma}\label{lem:gm_in_opt}
For every $i\in[d]$, there exist a point
$q_i\in\operatorname{conv}(C_i)$ and an open neighborhood
$U_i$ of $q_i$ on which $f^*$ is affine and satisfies
$f^*(c)-f^*(c_0)=z(c)$ for every $c\in U_i.
$
In particular, $\nabla f^*(c)=-M^{i-1}\one $ for every $c\in U_i.
$
\end{lemma}

\begin{proof}
Fix $i\in[d]$.
Consider the partition into convex cells, on each of which $f^*$ is convex, formed by the hinge lines of $f^*$ in $\mathbb{R}^2$.
By \Cref{fact:numregions}, these hinge lines form at most
$
q:=1+d+\binom d2
$ such convex cells, whose closures cover
$\mathbb R^2$. To avoid ambiguity, let us assign each candidate in $C_i$ to the closure of exactly one
cell containing it.
Since $|C_i|=3q$, the pigeonhole principle gives three
candidates $a_i^1,a_i^2,a_i^3\in C_i$ assigned to the
closure of the same cell $R_i$.
By construction of the instance, these three candidates are noncollinear and thus
lie in the interior of the same affine strip of $z$.
Let $\kappa:=f^*(c_0)$. We know that since $z$ realizes the unrestricted optimal candidate
rewards, every optimal MLP realizes those same rewards
up to an additive constant with
$ f^*(a_i^j)=z(a_i^j)+\kappa, (j=1,2,3).$
 
By continuity, the affine function defining $f^*$ on
$R_i$ takes the values $z(a_i^j)+\kappa$ at the three
candidates $a_i^j$.
An affine function on $\mathbb R^2$ is uniquely determined
by its values at three noncollinear points.
Thus this is the same affine function as the one defining
$z+\kappa$ on the $i$-th strip.
Because the three candidates are noncollinear, their
centroid $q_i:=\frac{a_i^1+a_i^2+a_i^3}{3}\in\operatorname{conv}(C_i)$
is an interior point of their triangle in $\overline{R_i}$, so $q_i$
lies in $R_i$.
Moreover, $q_i$ lies in the interior of the $i$-th
affine strip of $z$.
We can therefore choose an open neighborhood $U_i$
of $q_i$ contained in both $R_i$ and that strip such that
$
f^*(c)-f^*(c_0)=z(c),
$
for all $c\in U_i$. Since the gradient of $z$ on this strip is
$-M^{i-1}\one$, we also obtain
$
\nabla f^*(c)=-M^{i-1}\one
$ for all $ c\in U_i$.
\end{proof}
The next lemma shows that, for a suitable choice of \(M\), these gradients force all hinge lines to be parallel.

Let $\one:=(1,1)$. 
For $M>1$, define
$$
T_M
:=
\begin{pmatrix}
-\one^\top\\
-M\one^\top\\
\vdots\\
-M^{d-1}\one^\top
\end{pmatrix}
\in\mathbb R^{d\times2}.
$$

For the parameters $\beta,(\alpha_i,a_i,b_i,\eta_i)_{i=1}^d$
of $f^*$, define
$
u_i:=\alpha_i(a_i,b_i).
$
On each cell formed by the hinge lines, the gradient
of $f^*$ is the sum of those $u_i$ for which
$a_ix+b_iy+\eta_i>0$ on that cell.
Thus every gradient on such a cell is a subset sum
of $u_1,\dots,u_d$.

\begin{lemma}\label{lem:fstar_in_span}
There exists a finite set $\mathcal E_d\subset(1,\infty)$
such that, whenever $M\notin\mathcal E_d$, each vector
$u_i$ is a nonzero scalar multiple of $\one$.
Consequently, every hinge line has the form
$\varphi(c)=t$ for some $t\in\mathbb R$, and the
gradient of $f^*$ on every cell formed by these
lines is a scalar multiple of $\one$.
\end{lemma}

\begin{proof}
Let $T_M$ be the matrix whose $i$-th row is
$-M^{i-1}\one^\top$.
For each singular matrix $Y\in\{0,1\}^{d\times d}$,
choose a nonzero vector $w^Y\in\mathbb R^d$ satisfying
$(w^Y)^\top Y=0$, and define the nonzero polynomial
$P_Y(t):=\sum_{i=1}^d w_i^Y t^{i-1}$.
Let $\mathcal E_d$ contain all roots in $(1,\infty)$
of these polynomials.
This set is finite because there are finitely many
binary matrices and each polynomial has finitely many roots.

Fix $M\notin\mathcal E_d$.
No singular binary matrix $Y$ can satisfy $YV=T_M$
for any $V\in\mathbb R^{d\times2}$, since this would imply
$$
0=(w^Y)^\top YV
  =(w^Y)^\top T_M
  =-P_Y(M)\one^\top,
$$
contrary to the choice of $M$.

Now let $V$ have rows $u_1^\top,\dots,u_d^\top$.
By \Cref{lem:gm_in_opt}, each gradient
$-M^{i-1}\one$ is attained by $f^*$ and is therefore
a subset sum of $u_1,\dots,u_d$.
Recording these subsets as the rows of a binary matrix
$X$ gives $XV=T_M$.
The preceding argument shows that $X$ is invertible.
Thus $V=X^{-1}T_M$, so every row of $V$ is a scalar
multiple of $\one^\top$.

Moreover, no row of $V$ is zero.
Otherwise, replacing the corresponding column of $X$
with zeros would produce a singular binary matrix
$\widetilde X$ satisfying $\widetilde X V=XV=T_M$,
contradicting the same argument.
Hence every $u_i$ is a nonzero scalar multiple of $\one$.

Since $u_i=\alpha_i(a_i,b_i)\ne0$, the vector
$(a_i,b_i)$ is also a nonzero scalar multiple of $\one$.
Its hinge equation therefore has the form
$\varphi(c)=t$.
Finally, the gradient on each cell is a subset sum of
the $u_i$, so it is a scalar multiple of $\one$.
\end{proof}

For the remainder of the proof, fix
$M\in(1,\infty)\setminus\mathcal E_d$.

\begin{lemma}\label{lem:MLP-local}
In the degenerate instance, let
$p:=c_1=c_2,b:=\max_{c\in C_1}\varphi(c),$
and define the open strip
$U:=\{c\in\mathbb{R}^2:b<\varphi(c)<0\}.$
Then every optimal MLP $f^*$ of width at most $d$ satisfies
$$
f^*(c)-f^*(c_0)=z(c)=-\varphi(c)
\qquad\text{for every }c\in U.
$$
Moreover, no hinge line of $f^*$ intersects $U$.
\end{lemma}
\begin{proof}
Fix an optimal MLP $f^*$.
By Lemma~\ref{lem:fstar_in_span}, $f^*$ is constant on
each line $\varphi(c)=s$. Thus $f^*(c)-f^*(c_0)=h(\varphi(c))$
for a continuous piecewise-affine function $h:\mathbb{R}\rightarrow \mathbb{R}$ with
$h(0)=0$.
Since $h(s)=f^*(s/2,s/2)-f^*(c_0)$ and $f^*$ is an MLP, each MLP term contributes at most one breakpoint to $h$. Hence $h$ has at most $d$ breakpoints.

For each $i\in[d]$, take $q_i$ from
Lemma~\ref{lem:gm_in_opt} and set $\tau_i:=\varphi(q_i)$.
By the same lemma, it holds that $h=g$ near $\tau_i$, where the slope
is $-M^{i-1}$.
The ordering of the candidate blocks $C_i$ gives
$\tau_1>\tau_2>\cdots>\tau_d$ and $\tau_1\le b<0$.
Since these slopes are distinct, there is at least one
breakpoint between each consecutive pair of witnesses.
Thus at least $d-1$ breakpoints lie strictly to the left
of $\tau_1$, leaving at most one in $[\tau_1,0]$.

Near $\tau_1$, we have $h(s)=g(s)=-s$.
This equality extends throughout $[\tau_1,0]$:
if there is an interior breakpoint $t$, continuity
gives $h(t)=-t$, and the remaining affine piece joins
$(t,-t)$ to $(0,h(0))=(0,0)$, so it also equals $-s$.
Since $\tau_1\le b$ and $g(s)=-s$ on $[b,0]$,
we obtain $f^*(c)-f^*(c_0)=-\varphi(c)=z(c)$
throughout $U$. Finally, Lemma~\ref{lem:fstar_in_span} shows that each
term has a nonzero gradient vector $u_j$, and that
each hinge line has the form $\varphi(c)=t$.
The $d-1$ distinct breakpoints identified above require
at least $d-1$ different terms whose hinges lie outside
$U$.
Therefore at most one term can have a hinge intersecting
$U$.
Such a hinge would change the gradient, with no other
term available to cancel that change, contradicting
the fact that $f^*$ is affine on $U$.
Hence no hinge line intersects $U$.
\end{proof}

We now use Berge's maximum theorem to extend what we proved in \Cref{lem:MLP-local} to all sufficiently small perturbations of $c_1$. Since a degenerate optimal solution need not be unique, we
consider an open parameter neighborhood of the entire set of degenerate
optimal solutions. We will show that, for sufficiently small $\delta>0$,
every optimal $f^*$ satisfies $f^*(c_2)>f^*(c_1(\delta))$, contrary to
the unanimous voter ranking.

\begin{claim}\label{clm:MLPs-limit}
There exists a sufficiently small $\delta > 0$ such that any optimal solution of \eqref{eq:MLP-opt} satisfies $f^*(c_2) > f^*(c_1)$.
\end{claim}

Before proving this claim, let us see why this implies \Cref{thm:MLP-PO}. For $\delta$ as given by the claim, any optimal solution ranks $c_2$ over $c_1$. On the other hand, both voters $v_1$ and $v_2$ rank $c_1$ over $c_2$, so Pareto optimality is not satisfied.
\begin{proof}

Choose a closed ball $K$ around $p=c_2$ inside the
strip from \Cref{lem:MLP-local}.
At every degenerate optimal solution, no hinge line meets $K$
and the gradient there is $-\one$.
Since each hinge line lies a positive distance from $K$, compactness of $K$ implies that
sufficiently small parameter changes keep all hinge lines
outside $K$. The network therefore remains affine on
$K$, and its gradient varies continuously.
Thus there is an open parameter set $\mathcal O$
containing every degenerate optimal solution such that every
network represented in $\mathcal O$ is affine on $K$
and satisfies $\nabla f_\theta(p)\cdot(1,2)<0.$ Since the objective is continuous and $\Theta_d$ is
fixed and compact, Berge's theorem (\Cref{thm:berge}) implies that every
optimal solution for sufficiently small $\delta>0$ belongs
to $\mathcal O$.
Taking $\delta$ also small enough that $c_1(\delta)\in K$,
every such optimal solution satisfies
$$
f_\theta(c_1(\delta))-f_\theta(c_2)
=\delta\,\nabla f_\theta(p)\cdot(1,2)<0.
$$
\end{proof}

\section{Recovering Pareto Optimality with Uniform Data}\label{sec:sampling}
In this section, we return to linear reward functions. So far, we have considered axioms with respect to a fixed, finite candidate set, with voters' preferences over this set inducing a finite set of binary comparisons. In the linear social choice setting, however, binary comparisons have additional structure: a comparison is not only information about two candidates, but also information about a vector in the embedding space.

Indeed, a comparison between candidates \(a\) and \(b\) has an associated
difference vector \(b-a\). Its direction determines which voters prefer the
comparison, while its length \(\|b-a\|_2\) determines the scale at which the
comparison enters the loss. If voter \(i\) is represented by the linear reward
\(r_{v_i}(x)=\langle v_i,x\rangle\), then voter \(i\) prefers \(b\) to \(a\)
precisely when \(\langle v_i,b-a\rangle>0\). Thus, if voters are linear, we
can identify a binary comparison with its difference vector, and speak of
a voter preferring a direction.\footnote{This simple observation is useful
because it shows how quickly preferences generalize in the linear setting. If
voter \(i\) prefers the direction \(x\), then the voter also prefers every
positive multiple of \(x\). Moreover, if the voter prefers two directions \(x\)
and \(y\), then the voter also prefers any nonnegative combination of them,
since \(r_{v_i}(\lambda x+\mu y)=\langle v_i,\lambda x+\mu y\rangle
=\lambda\langle v_i,x\rangle+\mu\langle v_i,y\rangle>0\) for
\(\lambda,\mu\ge0\). Thus, once a few directions are known to be preferred, the
linearity assumption immediately determines preferences on the cone generated
by them.}

This identification lets us formulate a generalized version of PO, defined over
directions in \(\mathbb R^d\) rather than only over the finite set of observed
candidate comparisons. In particular, it applies also to directions for which no
comparison is explicitly present in the dataset.
\begin{definition}
We say that $\theta \in \mathbb{R}^d$ is Pareto optimal (PO) over $\mathbb{R}^d$ with respect to voters $v_1, \dots, v_n \in \mathbb{R}^d \setminus \{0\}$ if whenever a direction $x \in S^{d-1}$ satisfies $\inner{v_i}{x} > 0$ for all $i$, we also have $\inner{\theta}{x} > 0$.
\end{definition}
Representing comparisons by their difference vectors, we describe
comparison data by a measure \(\mu\) over directions
\(x\in S^{d-1}\), recording their weights, and a length function
\(R:S^{d-1}\to(0,\infty)\). This representation assumes that
comparisons in the same direction have the same length. This condition can be enforced by normalizing each difference vector \(b-a\) to \((b-a)/\|b-a\|_2\) before evaluating the loss, giving \(R(x)=1\) for every direction. This preserves the linear voters’ pairwise preferences while changing how the comparisons enter the training objective.
Finite datasets satisfying this condition correspond to discrete
measures \(\mu\), with point masses at the observed directions.
Here, we focus on the continuous case, where
\(d\mu(x)=\rho(x)\,dx\).

For this continuous version, the total loss takes the form
\begin{equation}\label{eq:loss-general-directions}
\totalloss(\theta)
:=
\sum_{i=1}^n
\int_{x\in S^{d-1},\,\inner{v_i}{x}\ge0}
\ell\bigl(-\inner{\theta}{R(x)\cdot x}\bigr)\rho(x)\,dx, 
\end{equation}
where \(\rho\) is a density function and \(v_1,\dots,v_n\in\mathbb R^d\setminus\{0\}\) are the voter directions,
inducing rankings according to the linear rewards
\(r_{v_i}(x)=\inner{v_i}{x}\).

The following proposition shows that uniform comparison directions do not
suffice to guarantee PO when comparison lengths vary.

\begin{restatable}[\deferredsym]{proposition}{ellipsefailsPO}\label{prop:ellipse-fails-po}
The loss-based voting rule with BT loss fails PO even when
comparison directions are uniformly distributed. This already holds in two
dimensions with two weighted voters, when comparison lengths are given by
the radial function of a centered, axis-aligned ellipse.
\end{restatable}

\subsection{Pareto Optimality}\label{subsec:uniform-PO}

The main positive result of this section concerns the most symmetric special
case of this model. We consider an \emph{idealized data setting} in which
comparisons cover all possible directions uniformly and every comparison has
unit length.\footnote{The conclusion of \Cref{thm:limit-PO} also holds for
a common bounded distribution of positive comparison lengths \(T\),
independent of direction and voter. The averaged loss
\(\widetilde\ell(s)=\mathbb E[\ell(Ts)]\) satisfies the same assumptions,
with \(\widetilde\ell'(0)=\mathbb E[T]\ell'(0)>0\).}
Equivalently, \(R(x)=1\) and \(\rho(x)\) is constant, so the total
loss becomes
\begin{equation}\label{eq:loss-limit}
\totalloss(\theta)
:=
\sum_{i=1}^n
\int_{x\in S^{d-1},\,\inner{v_i}{x}\ge0}
\ell(-\inner{\theta}{x})\,dx .
\end{equation}

The following theorem establishes PO for a broad class of loss-based rules in this idealized setting.
\begin{theorem}\label{thm:limit-PO}
In the idealized data setting, any loss-based voting rule with a loss function $\ell$ that is convex, nondecreasing, lower bounded, and differentiable with $\ell'(0) > 0$ satisfies PO, provided that at least two voters have distinct directions.
\end{theorem}

\begin{proof}[Proof of \Cref{thm:limit-PO}]
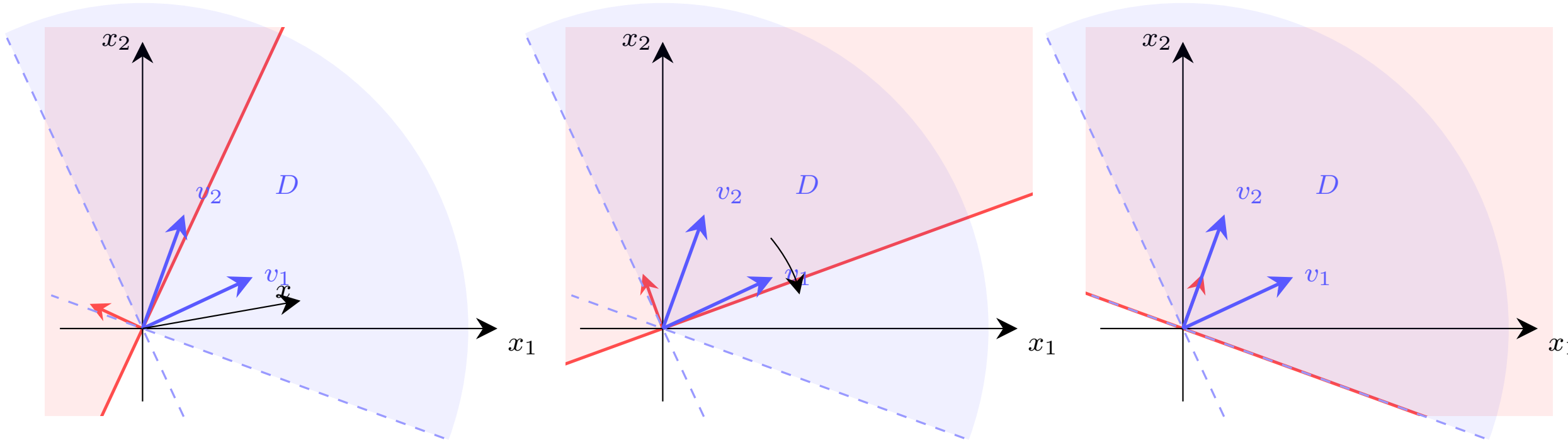
\begin{figure}[t]
  \centering
  \resizebox{\textwidth}{!}{\begin{tikzpicture}[scale=1.0, >=Stealth]
\scriptsize

\def\W{4.8}
\def\H{4.0}
\def\Gap{0.55}
\def\Ox{1.00}
\def\Oy{0.90}

\def\va{25}    %
\def\vb{70}    %

\def\edgeA{-20}
\def\edgeB{115}
\def\Rcone{3.35}

\def\thone{65}
\def\thtwo{20}
\def\ththree{\edgeA}

\tikzset{
  axis/.style      = {thin, black, -{Stealth[length=2.2mm,width=2.2mm]}},
  voter/.style     = {line width=1.0pt, blue!65, -{Stealth[length=2.4mm,width=2.4mm]}},
  vboundary/.style = {blue!40, dashed, line width=0.6pt},
  hplane/.style    = {fill=red!8, draw=none},
  hline/.style     = {line width=0.9pt, red!70},
  hnormal/.style   = {line width=0.9pt, red!70, -{Stealth[length=2.0mm,width=2.0mm]}},
  cone/.style      = {fill=blue, fill opacity=0.06, draw=none},
  labelblue/.style = {text=blue!65}
}

\begin{scope}[shift={(0,0)}]
  \coordinate (O) at (\Ox,\Oy);

  \begin{scope}
    \clip (0,0) rectangle (\W,\H);
    \begin{scope}[shift={(O)}, rotate=\thone]
      \fill[hplane] (-5,0) rectangle (5,5);
      \draw[hline] (-5,0) -- (5,0);
      \draw[hnormal] (0,0) -- ++(90:0.60);
    \end{scope}
  \end{scope}

  \draw[axis] (0.15,\Oy) -- (\W-0.15,\Oy) node[below right] {$x_1$};
  \draw[axis] (\Ox,0.15) -- (\Ox,\H-0.15) node[left] {$x_2$};

  \fill[cone] (O) -- ++(\edgeA:\Rcone)
    arc[start angle=\edgeA,end angle=\edgeB,radius=\Rcone] -- cycle;

  \begin{scope}[shift={(O)}, rotate={\va+90}]
    \draw[vboundary] (-1,0) -- (3.3,0);
  \end{scope}
  \begin{scope}[shift={(O)}, rotate={\vb+90}]
    \draw[vboundary] (-3.3,0) -- (1,0);
  \end{scope}

  \draw[voter] (O) -- ++(\va:1.25)
    node[anchor=west, labelblue] {$v_1$};
  \draw[voter] (O) -- ++(\vb:1.25)
    node[anchor=south west, labelblue] {$v_2$};

  \node[labelblue] at ($(O)+(45:2.10)$) {$D$};

  \draw[black, -{Stealth[length=2.2mm,width=2.2mm]}] (O) -- ++(10:1.65);
  \node at ($(O)+(1.45,0.38)$) {$x$};
\end{scope}

\begin{scope}[shift={(\W+\Gap,0)}]
  \coordinate (O) at (\Ox,\Oy);

  \begin{scope}
    \clip (0,0) rectangle (\W,\H);
    \begin{scope}[shift={(O)}, rotate=\thtwo]
      \fill[hplane] (-5,0) rectangle (5,5);
      \draw[hline] (-5,0) -- (5,0);
      \draw[hnormal] (0,0) -- ++(90:0.60);
    \end{scope}
  \end{scope}

  \draw[axis] (0.15,\Oy) -- (\W-0.15,\Oy) node[below right] {$x_1$};
  \draw[axis] (\Ox,0.15) -- (\Ox,\H-0.15) node[left] {$x_2$};

  \fill[cone] (O) -- ++(\edgeA:\Rcone)
    arc[start angle=\edgeA,end angle=\edgeB,radius=\Rcone] -- cycle;

  \begin{scope}[shift={(O)}, rotate={\va+90}]
    \draw[vboundary] (-1,0) -- (3.3,0);
  \end{scope}
  \begin{scope}[shift={(O)}, rotate={\vb+90}]
    \draw[vboundary] (-3.3,0) -- (1,0);
  \end{scope}

  \draw[voter] (O) -- ++(\va:1.25)
    node[anchor=west, labelblue] {$v_1$};
  \draw[voter] (O) -- ++(\vb:1.25)
    node[anchor=south west, labelblue] {$v_2$};

  \node[labelblue] at ($(O)+(45:2.10)$) {$D$};

  \draw[black, -{Stealth[length=2.0mm,width=2.0mm]}]
    ($(O)+(40:1.45)$)
    arc[start angle=40,end angle=14,radius=1.45];
\end{scope}

\begin{scope}[shift={(2*\W+2*\Gap,0)}]
  \coordinate (O) at (\Ox,\Oy);

  \begin{scope}
    \clip (0,0) rectangle (\W,\H);
    \begin{scope}[shift={(O)}, rotate=\ththree]
      \fill[hplane] (-5,0) rectangle (5,5);
      \draw[hline] (-5,0) -- (5,0);
      \draw[hnormal] (0,0) -- ++(90:0.60);
    \end{scope}
  \end{scope}

  \draw[axis] (0.15,\Oy) -- (\W-0.15,\Oy) node[below right] {$x_1$};
  \draw[axis] (\Ox,0.15) -- (\Ox,\H-0.15) node[left] {$x_2$};

  \fill[cone] (O) -- ++(\edgeA:\Rcone)
    arc[start angle=\edgeA,end angle=\edgeB,radius=\Rcone] -- cycle;

  \begin{scope}[shift={(O)}, rotate={\va+90}]
    \draw[vboundary] (-1,0) -- (3.3,0);
  \end{scope}
  \begin{scope}[shift={(O)}, rotate={\vb+90}]
    \draw[vboundary] (-3.3,0) -- (1,0);
  \end{scope}

  \draw[voter] (O) -- ++(\va:1.25)
    node[anchor=west, labelblue] {$v_1$};
  \draw[voter] (O) -- ++(\vb:1.25)
    node[anchor=south west, labelblue] {$v_2$};

  \node[labelblue] at ($(O)+(45:2.10)$) {$D$};
\end{scope}

\end{tikzpicture}}
  \caption{Geometric intuition for the proof of \Cref{thm:limit-PO}.
The blue rays indicate voter directions, and the blue cone \(P\) is the intersection of the voter half-planes. The red line is the boundary of the half-plane \(H_\theta\). If \(H_\theta\) fails to contain \(P\), the proof can be understood as rotating this half-plane toward \(P\) until it reaches a supporting position.}
  \label{fig:rotating-halfplane}
\end{figure}

Let $\totalloss_i$ denote the loss with respect to voter $i$, i.e.,
$$\totalloss_i(\theta) := \int_{x \in S^{d-1}, \inner{v_i}{x} \geq 0} \ell(-\inner{\theta}{x}) \, dx = \frac{1}{2} \int_{x \in S^{d-1}} \ell(-\sgn(\inner{v_i}{x}) \cdot \inner{\theta}{x}) \, dx,$$
where the sign function is defined as
\[
\sgn(t)=1 \text{ if } t>0,\quad
\sgn(t)=0 \text{ if } t=0,\quad
\sgn(t)=-1 \text{ if } t<0 .
\]

Then we can write $\totalloss(\theta) = \sum_{i=1}^n \totalloss_i(\theta)$. 

\begin{restatable}[\deferredsym]{claim}{limitcloserbetter}\label{clm:limit-closer-better}
If $\theta,\theta'\in\mathbb{R}^d$ satisfy $\|\theta\|_2=\|\theta'\|_2$ and $\langle \theta',v_i\rangle>\langle \theta,v_i\rangle$, then $\mathcal{L}_i(\theta')<\mathcal{L}_i(\theta)$.
\end{restatable}

\begin{restatable}[\deferredsym]{claim}{limitminattained}\label{clm:limit-min-attained}
If at least two voters have distinct directions, then $\mathcal{L}$ attains its minimum.
\end{restatable}

\begin{restatable}[\deferredsym]{claim}{limitnotzero}\label{clm:limit-not-zero}
If there exists $x \in S^{d-1}$ such that $\inner{v_i}{x} > 0$ for all voters $i \in [n]$, then $\theta = 0$ is not a minimum of $\totalloss$.
\end{restatable}

Finally, we use these three claims to prove the following.

\begin{claim}\label{clm:limit-minimizer-PO}
Any minimizer $\theta^*$ of $\totalloss$ satisfies PO over $\mathbb{R}^d$ with respect to the voters $v_1, \dots, v_n$.
\end{claim}
\begin{proof}
First of all, note that if there does not exist $x \in S^{d-1}$ such that $\inner{v_i}{x} > 0$ for all $i \in [n]$, then PO is trivially satisfied. Thus, from now on assume that the set $D := \{x \in S^{d-1}: \inner{v_i}{x} > 0 \text{ for all } i \in [n]\}$ is not empty. In particular, by \Cref{clm:limit-not-zero}, this implies any minimizer $\theta^*$ of $\totalloss$ is not the zero vector.

Consider any $\theta \neq 0$ such that there exists $x \in D$ such that $\inner{\theta}{x} \leq 0$. We will show that there exists $\theta'$ with $\totalloss(\theta') < \totalloss(\theta)$ and thus $\theta$ is not optimal. Since the set $D$ is open, we can assume without loss of generality that in fact $\inner{\theta}{x} < 0$. Furthermore, for simplicity we will assume that $\|\theta\|_2 = 1 = \|x\|_2$. If that is not the case, then one can simply scale $x$ so that it has the same length as $\theta$ and the same proof idea applies.

Let $\theta' := (1+\delta)\theta + \eps x$. We will show that for some carefully selected $\eps, \delta > 0$, $\theta'$ satisfies $\inner{\theta'}{v_i} > \inner{\theta}{v_i}$ and $\|\theta'\|_2 = \|\theta\|_2 = 1$. As a result, using \Cref{clm:limit-closer-better}, it will follow that $\totalloss_i(\theta') < \totalloss_i(\theta)$ for all $i \in [n]$, and thus $\totalloss(\theta') < \totalloss(\theta)$, as desired.

The condition $\inner{\theta'}{v_i} > \inner{\theta}{v_i}$ holds as long as $\delta \inner{v_i}{\theta} + \eps \inner{v_i}{x} > 0$, which holds as long as
$\eps > -\frac{\inner{v_i}{\theta}}{\inner{v_i}{x}}\delta$
holds for all $i \in [n]$. Let $M := 2 + \max\{0, -1/\inner{\theta}{x}, \max_i -\inner{v_i}{\theta}/\inner{v_i}{x}\}$. Then letting $\eps := M\delta$ ensures that the aforementioned condition is satisfied.

It remains to pick $\delta > 0$ such that $\|\theta'\|_2 = 1$. We can write
\begin{equation*}
\begin{split}
\|\theta'\|_2^2 = \inner{\theta'}{\theta'} &= \inner{(1+\delta)\theta + M\delta x}{(1+\delta)\theta + M\delta x}\\
&= (1+\delta)^2 + 2(1+\delta)M\delta\inner{\theta}{x} + M^2\delta^2\\
&= 1 + 2\delta + \delta^2 +2M\delta \inner{\theta}{x} + 2\delta^2M\inner{\theta}{x} + M^2 \delta^2\\
&= 1 + \delta^2(1 + 2M\inner{\theta}{x} + M^2) + \delta(2 + 2M\inner{\theta}{x}).
\end{split}
\end{equation*}
Note that $1 + 2M\inner{\theta}{x} + M^2 \geq 1 - 2M + M^2 > 0$ and $2 + 2M\inner{\theta}{x} < 0$, by the choice of $M$ and since $\inner{\theta}{x} \in [-1,0)$. As a result, we have that for sufficiently large $\delta > 0$, $\|\theta'\|_2 > 1$, and for sufficiently small $\delta > 0$, $\|\theta'\|_2 < 1$. By the intermediate value theorem, if follows that there exists $\delta > 0$ such that $\|\theta'\|_2 = 1$, and this completes the proof.
\end{proof}

\Cref{thm:limit-PO} follows by combining \Cref{clm:limit-min-attained}, which guarantees that a minimizer exists, with \Cref{clm:limit-minimizer-PO}, which guarantees that any minimizer satisfies PO.

\end{proof}

\subsection{Pairwise-Majority-Consistency}

On the other hand, we can show that the analogous definition of Pairwise Majority Consistency (PMC) is not satisfied even in this setting.

\begin{definition}
We say that $\theta \in \mathbb{R}^d$ is pairwise majority consistent (PMC) over $\mathbb{R}^d$ with respect to voters $v_1, \dots, v_n \in \mathbb{R}^d \setminus \{0\}$ if whenever a direction $x \in S^{d-1}$ satisfies $\inner{v_i}{x} > 0$ for a strict majority of all $i$, we also have $\inner{\theta}{x} > 0$. A voting rule satisfies PMC if it outputs a PMC vector whenever one exists.
\end{definition}

\begin{restatable}[\deferredsym]{theorem}{noPMC}\label{thm:limit-no-PMC}
In the idealized data setting, any loss-based voting rule with a loss function $\ell$ that is strictly convex, nondecreasing, lower bounded, and differentiable with $\ell'(0) > 0$ fails PMC.
\end{restatable}

Although PMC fails in this uniform data setting, it is not at all obvious whether PMC is even desirable here. For example, if a $p$ fraction of votes come from direction $v_1$ and a $(1-p)$ fraction from direction $v_2$, then in the uniform data setting PMC becomes a discontinuous requirement: the rule must output direction $v_1$ whenever $p>0.5$ and direction $v_2$ whenever $p<0.5$. In many applications, one may instead prefer to interpolate between the two. See the work by~\cite{LedererPW2024-squared-Kemeny} for a recent discussion of similar considerations in rank aggregation (from which PMC is inherited).

\section{Discussion and Future Directions}
We contribute to the recent effort to bring the axiomatic approach of social choice into AI alignment. Much of classical social choice studies rankings over a fixed set of alternatives. Reward models, however, must assign scores through a restricted model class and are used to compare responses beyond their training data. If social choice is to inform alignment, its axioms and, more generally, its theoretical models must more closely reflect this setting.

In the linear social choice setting, \citet{ge2024axioms} showed that an optimally trained linear reward model can violate Pareto optimality (PO), reversing a unanimous preference. In response, we investigated whether PO can be recovered while retaining loss-based training by (1) increasing the capacity of the reward model or (2) improving the comparison data.
The first point acknowledges that, while many analyses assume unconstrained rewards, such optima are unlikely to be realized in practice. For (1), \Cref{thm:poly-PO,thm:MLP-PO} establish that PO can fail both for polynomial rewards of bounded degree and for ReLU MLPs with one hidden layer. 
We conjecture that our impossibility results extend to MLPs of arbitrary fixed depth and width over frozen embeddings, and beyond that to transformers of fixed architecture, under the same loss assumptions and suitable finite parameter bounds.

As a step toward generalization, we show that in structured settings (e.g., linear social choice), axioms specified over a fixed alternative set admit natural analogues over the entire embedding space. %
For (2), we establish PO for linear rewards in an idealized setting with uniformly distributed comparison directions and equal comparison lengths. Examining normalized embedding differences—for example, through PCA—can help identify directional imbalance.
While the lengths of comparison directions can be balanced by renormalizing the loss, achieving more uniform directional coverage may require reweighting comparisons, and in the worst case, collecting additional data. A natural next step is to derive explicit sampling bounds in the uniform setting. Beyond random sampling, a promising direction is to formalize reweighting and data-query strategies—can we gather data to attain PO in a more targeted (and more efficient) manner?
While proposing new axioms is not our focus, future research should consider how other axioms extend to embedding spaces and can be satisfied through data selection.

Our results raise a broader question about the role of comparison-data distributions in axiomatic guarantees. How much should these guarantees depend on which comparisons occur frequently, and does the underlying embedding space admit a meaningful reference measure against which to assess or rebalance the data? Understanding this distinction would help determine when changing the comparison distribution is justified, and when the guarantees themselves should reflect the distribution relevant to the application.
\section*{Acknowledgments}

Sonja Kraiczy was supported by the PIBBSS Fellowship (Principles of Intelligence) and the Cooperative AI Foundation.

\section*{AI Usage}
As part of our revisions since the ICLR 2026 submission, we used GPT-6 to complete the proof of an illustrative counterexample, now included as Proposition 1. We also used it to improve the presentation and correct minor issues in the introductory examples in Section 3. The authors have verified the proof and take full responsibility for the paper's content.

\bibliography{references}

@article{ge2024axioms,
  title={{Axioms for AI Alignment from Human Feedback}},
  author={Ge, Luise and Halpern, Daniel and Micha, Evi and Procaccia, Ariel D. and Shapira, Itai and Vorobeychik, Yevgeniy and Wu, Junlin},
  journal={Advances in Neural Information Processing Systems},
  volume={37},
  pages={80439--80465},
  year={2024}
}

@inproceedings{park2023linear,
  title={The Linear Representation Hypothesis and the Geometry of Large Language Models},
  author={Park, Kiho and Choe, Yo Joong and Veitch, Victor},
  booktitle={Forty-first International Conference on Machine Learning},
  year={2023}
}

@book{berge1997topological,
  title={Topological Spaces: including a treatment of multi-valued functions, vector spaces, and convexity},
  author={Berge, Claude},
  year={1997},
  publisher={Courier Corporation}
}

@inproceedings{conitzer2024social,
  title={Position: social choice should guide {AI} alignment in dealing with diverse human feedback},
  author={Conitzer, Vincent and Freedman, Rachel and Heitzig, Jobst and Holliday, Wesley H and Jacobs, Bob M and Lambert, Nathan and Moss{\'e}, Milan and Pacuit, Eric and Russell, Stuart and Schoelkopf, Hailey and others},
  booktitle={Proceedings of the 41st International Conference on Machine Learning},
  pages={9346--9360},
  year={2024}
}

@article{procaccia2025clone,
  title={Clone-robust {AI} alignment},
  author={Procaccia, Ariel D and Schiffer, Benjamin and Zhang, Shirley},
  journal={arXiv preprint arXiv:2501.09254},
  year={2025}
}

@article{bai2022constitutional,
  title={Constitutional {AI}: Harmlessness from {AI} feedback},
  author={Bai, Yuntao and Kadavath, Saurav and Kundu, Sandipan and Askell, Amanda and Kernion, Jackson and Jones, Andy and Chen, Anna and Goldie, Anna and Mirhoseini, Azalia and McKinnon, Cameron and others},
  journal={arXiv preprint arXiv:2212.08073},
  year={2022}
}

@article{bai2022training,
  title={Training a helpful and harmless assistant with reinforcement learning from human feedback},
  author={Bai, Yuntao and Jones, Andy and Ndousse, Kamal and Askell, Amanda and Chen, Anna and DasSarma, Nova and Drain, Dawn and Fort, Stanislav and Ganguli, Deep and Henighan, Tom and others},
  journal={arXiv preprint arXiv:2204.05862},
  year={2022}
}

@book{arrow1951social, author = {Arrow, Kenneth J.}, title = {Social Choice and Individual Values}, publisher = {Wiley}, year = {1951}}

@article{golz2025distortion,
  title={Distortion of {AI} Alignment: Does Preference Optimization Optimize for Preferences?},
  author={G{\"o}lz, Paul and Haghtalab, Nika and Yang, Kunhe},
  journal={arXiv preprint arXiv:2505.23749},
  year={2025}
}

@inproceedings{dai2024mapping,
  title={Mapping Social Choice Theory to {RLHF}},
  author={Dai, Jessica and Fleisig, Eve},
  booktitle={ICLR 2024 Workshop on Reliable and Responsible Foundation Models},
  year={2024}
}

@article{elkind2022preference,
  title={Preference restrictions in computational social choice: A survey},
  author={Elkind, Edith and Lackner, Martin and Peters, Dominik},
  journal={arXiv preprint arXiv:2205.09092},
  year={2022}
}

@book{zaslavsky1975facing,
  title={Facing up to arrangements: Face-count formulas for partitions of space by hyperplanes: Face-count formulas for partitions of space by hyperplanes},
  author={Zaslavsky, Thomas},
  volume={154},
  year={1975},
  publisher={American Mathematical Soc.}
}

@inproceedings{munos2023nash,
  title={Nash Learning from Human Feedback},
  author={Munos, Remi and Valko, Michal and Calandriello, Daniele and Azar, Mohammad Gheshlaghi and Rowland, Mark and Guo, Zhaohan Daniel and Tang, Yunhao and Geist, Matthieu and Mesnard, Thomas and Fiegel, C{\^o}me and others},
  booktitle={International Conference on Machine Learning},
  pages={36743--36768},
  year={2024},
  organization={PMLR}
}

@inproceedings{procaccia2006distortion,
  title={The distortion of cardinal preferences in voting},
  author={Procaccia, Ariel D and Rosenschein, Jeffrey S},
  booktitle={International Workshop on Cooperative Information Agents},
  pages={317--331},
  year={2006},
  organization={Springer}
}

@article{fishburn1984probabilistic,
  title={Probabilistic social choice based on simple voting comparisons},
  author={Fishburn, Peter C},
  journal={The Review of Economic Studies},
  volume={51},
  number={4},
  pages={683--692},
  year={1984},
  publisher={Wiley-Blackwell}
}

@inproceedings{siththaranjan2023distributional,
  author    = {Anand Siththaranjan and Cassidy Laidlaw and Dylan Hadfield-Menell},
  title     = {Distributional Preference Learning: Understanding and Accounting for Hidden Context in {RLHF}},
  booktitle = {The Twelfth International Conference on Learning Representations (ICLR)},
  year      = {2024}
}

@article{christiano2017deep,
  title={Deep reinforcement learning from human preferences},
  author={Christiano, Paul F and Leike, Jan and Brown, Tom and Martic, Miljan and Legg, Shane and Amodei, Dario},
  journal={Advances in neural information processing systems},
  volume={30},
  year={2017}
}

@inproceedings{cui2023ultrafeedback,
  title={ULTRAFEEDBACK: Boosting Language Models with Scaled {AI} Feedback},
  author={Cui, Ganqu and Yuan, Lifan and Ding, Ning and Yao, Guanming and He, Bingxiang and Zhu, Wei and Ni, Yuan and Xie, Guotong and Xie, Ruobing and Lin, Yankai and others},
  booktitle={Forty-first International Conference on Machine Learning},
  year = {2023}
}

@article{glaese2022improving,
  title={Improving alignment of dialogue agents via targeted human judgements},
  author={Glaese, Amelia and McAleese, Nat and Tr{\k{e}}bacz, Maja and Aslanides, John and Firoiu, Vlad and Ewalds, Timo and Rauh, Maribeth and Weidinger, Laura and Chadwick, Martin and Thacker, Phoebe and others},
  journal={arXiv preprint arXiv:2209.14375},
  year={2022}
}

@article{sun2023principle,
  title={Principle-driven self-alignment of language models from scratch with minimal human supervision},
  author={Sun, Zhiqing and Shen, Yikang and Zhou, Qinhong and Zhang, Hongxin and Chen, Zhenfang and Cox, David and Yang, Yiming and Gan, Chuang},
  journal={Advances in Neural Information Processing Systems},
  volume={36},
  pages={2511--2565},
  year={2023}
}

@book{arrow2010handbook,
  title={Handbook of social choice and welfare},
  author={Arrow, Kenneth J and Sen, Amartya and Suzumura, Kotaro},
  volume={2},
  year={2010},
  publisher={Elsevier}
}

@book{brandt2016handbook,
  title={Handbook of computational social choice},
  author={Brandt, Felix and Conitzer, Vincent and Endriss, Ulle and Lang, J{\'e}r{\^o}me and Procaccia, Ariel D},
  year={2016},
  publisher={Cambridge University Press}
}

@article{noothigattu2020axioms,
  title={Axioms for learning from pairwise comparisons},
  author={Noothigattu, Ritesh and Peters, Dominik and Procaccia, Ariel D},
  journal={Advances in Neural Information Processing Systems},
  volume={33},
  pages={17745--17754},
  year={2020}
}

@misc{maurarivero2025,
      title={Jackpot! Alignment as a Maximal Lottery}, 
      author={Roberto-Rafael Maura-Rivero and Marc Lanctot and Francesco Visin and Kate Larson},
      year={2025},
      eprint={2501.19266},
      archivePrefix={arXiv},
      primaryClass={cs.AI},
      url={https://arxiv.org/abs/2501.19266}, 
}

@inproceedings{Dudik2015,
  author    = {Miroslav Dud{\'i}k and Katja Hofmann and Robert E. Schapire
               and Aleksandrs Slivkins and Masrour Zoghi},
  title     = {Contextual Dueling Bandits},
  booktitle = {Proceedings of the 28th Conference on Learning Theory},
  pages     = {563--587},
  year      = {2015}
}

@inproceedings{WangLiuJin2023,
  author    = {Yuanhao Wang and Qinghua Liu and Chi Jin},
  title     = {Is {RLHF} More Difficult than Standard {RL}? A Theoretical Perspective},
  booktitle = {Advances in Neural Information Processing Systems 36 (NeurIPS 2023)},
  year      = {2023}
}

@inproceedings{LedererPW2024-squared-Kemeny,
author = {Lederer, Patrick and Peters, Dominik and W\k{a}s, Tomasz},
title = {The Squared {Kemeny} Rule for Averaging Rankings},
year = {2024},
doi = {10.1145/3670865.3673521},
booktitle = {Proceedings of the 25th ACM Conference on Economics and Computation (EC)},
pages = {755},
numpages = {1},
}

@article{Anderson2009BordaZermelo,
  title   = {Relationships between Borda voting and Zermelo ranking},
  author  = {Anderson, Lowell Bruce and Dandurova, Helena and Falk, James E. and Yeganova, Lana},
  journal = {Social Choice and Welfare},
  year    = {2009},
  volume  = {32},
  number  = {3},
  pages   = {355--365},
  doi     = {10.1007/s00355-008-0327-7},
  publisher = {Springer}
}

@article{qiu2024representative,
  title={Representative social choice: From learning theory to {AI} alignment},
  author={Qiu, Tianyi},
  journal={arXiv preprint arXiv:2410.23953},
  year={2024}
}

@article{xiao2025theoretical,
  title={Theoretical Tensions in {RLHF}: Reconciling Empirical Success with Inconsistencies in Social Choice Theory},
  author={Xiao, Jiancong and Shi, Zhekun and Liu, Kaizhao and Long, Qi and Su, Weijie J},
  journal={arXiv preprint arXiv:2506.12350},
  year={2025}
}

@article{kim2025population,
  title={Population-Proportional Preference Learning from Human Feedback: An Axiomatic Approach},
  author={Kim, Kihyun and Zhang, Jiawei and Ozdaglar, Asuman and Parrilo, Pablo A},
  journal={arXiv preprint arXiv:2506.05619},
  year={2025}
}

@article{bradley1952rank,
  title   = {Rank Analysis of Incomplete Block Designs. I. The Method of Paired Comparisons},
  author  = {Bradley, Ralph Allan and Terry, Milton E.},
  journal = {Biometrika},
  volume  = {39},
  number  = {3-4},
  pages   = {324--345},
  year    = {1952},
  doi     = {10.1093/biomet/39.3-4.324}
}

@article{zermelo1929turnier,
  title   = {{Die Berechnung der Turnier-Ergebnisse als ein Maximumproblem der Wahrscheinlichkeitsrechnung}},
  author  = {Zermelo, Ernst},
  journal = {Mathematische Zeitschrift},
  volume  = {29},
  pages   = {436--460},
  year    = {1929},
  doi     = {10.1007/BF01180541}
}

@book{Rockafellar1970-convex-analysis,
  author = {Rockafellar, R. Tyrrell},
  publisher = {Princeton University Press},
  series = {Princeton Mathematical Series},
  title = {Convex analysis},
  year = {1970}
}

@book{lee2013smooth,
  author={Lee, John M.},
  title={Introduction to Smooth Manifolds},
  edition={2},
  series={Graduate Texts in Mathematics},
  volume={218},
  year={2013},
  publisher={Springer},
  address={New York},
  doi={10.1007/978-1-4419-9982-5},
  url={https://doi.org/10.1007/978-1-4419-9982-5}
}
\appendix

\section{Additional Related Work}
As an alternative to unguided preference elicitation (“which response do you prefer?”), several authors \citep{bai2022training,glaese2022improving} proposed principle-guided supervision: annotators judge responses with respect to explicit criteria such as helpfulness, honesty, and harmlessness (HHH). In current practice, RLHF often elicits binary preferences from users as a general signal without specifying how to evaluate “better,” and human comparisons are noisy—commonly modeled with the Bradley–Terry framework. Asking raters to compare responses with respect to a named principle (e.g., “which is more harmful?”) provides a clearer target and greater control, reducing variance from idiosyncratic tastes (e.g., preferring a flattering style). Taking this perspective, we can view the principles as the voters. As models become more capable, human raters can be replaced or augmented with model-based judges, yielding Reinforcement Learning from AI Feedback (RLAIF). This makes alignment methods more scalable. E.g., it is cheap to score the same pairwise comparison against multiple principles, which may legitimately disagree (a response can be more helpful yet more harmful). Constitutional AI (CAI) \citep{bai2022constitutional} is a concrete framework built around RLAIF: it introduces an explicit “constitution”—a curated set of principles— which judges which of two responses better adheres to a given principle. Notably, when CAI was introduced, Anthropic reported a Pareto improvement on the helpfulness–harmlessness frontier: increased harmlessness at fixed helpfulness relative to RLHF baselines.
Many other works implement constitutional-style alignment, often under different labels, by conditioning binary comparisons on explicit criteria (rubrics/attributes), using verifier- or judge-guided feedback, or training separate objectives that are later combined \citep{cui2023ultrafeedback, sun2023principle,glaese2022improving}.

\section{Missing Proofs from \Cref{sec:genrewards}}
\uniquerewards*

\begin{proof}
Recall that $r_1 = 0$. For each $j \in \{2, \dots, m\}$, we have that both the terms $(1-\alpha) \cdot \ell(r_j)$ and $\alpha \cdot \ell(-r_j)$ appear in $\mathcal{L}$. Since $\ell$ is convex and $\ell'(0) > 0$, it follows that $\lim_{x \to +\infty} \ell(x) = +\infty$. Furthermore, recall that $\ell$ is lower bounded over $\mathbb{R}$. As a result, it follows that $\mathcal{L}(r) \to +\infty$ when $|r_j| \to +\infty$.

Since the loss function $\ell$ is strictly convex, the function $\phi: (r_2, \dots, r_m) \mapsto \mathcal{L}(0,0,r_2,\dots,r_m)$ is also strictly convex.\footnote{Here we use the fact that for $(r_2, \dots, r_m) \neq (r_2', \dots, r_m')$, there exists $j \in \{2, \dots, m\}$ such that $r_j \neq r_j'$, and thus $\mathcal{L}$ contains a term for which $\ell((r_j+r_j')/2) < \ell(r_j)/2 + \ell(r_j')/2$.} Thus, together with the previous paragraph, it follows that $\phi$ attains its unique minimum $r^*$.

It remains to argue that the optimal rewards are ordered $0 = r_1^* < r_2^* < \dots < r_m^*$. To prove this, we first introduce some additional notation. Define $h: \mathbb{R} \to \mathbb{R}$, $x \mapsto \alpha \cdot \ell(x) + (1-\alpha) \cdot \ell(-x)$. Note that $h$ is convex and differentiable, and thus its derivative $h'$ is nondecreasing. Furthermore, it satisfies
$$h'(0) = \alpha \cdot \ell'(0) - (1-\alpha) \cdot \ell'(0) = (2\alpha - 1) \cdot \ell'(0) > 0$$
since $\alpha > 1/2$ and $\ell'(0)$. We can rewrite the total loss function as
$$\mathcal{L}(0,0,r_2,\dots,r_m) = \ell(0) + 2\sum_{j=2}^m h(-r_j) + \sum_{i=2}^{m-1} \sum_{j=i+1}^m h(r_i - r_j)$$
and thus the partial derivatives for all $i \geq 2$ as
$$\frac{\partial \mathcal{L}}{\partial r_{i}}(r) = -2 h'(-r_i) - \sum_{k=2}^{i-1} h'(r_k - r_i) + \sum_{k=i+1}^{m} h'(r_i - r_k).$$

Assume towards a contradiction that $r_i^* \geq r_{i+1}^*$ for some $i \in \{2, \dots, m-1\}$. We will handle the remaining case $i=1$ separately at the end. Since $r^*$ is the optimal solution and $\mathcal{L}$ is differentiable, we have $\partial \mathcal{L}/\partial r_i(r^*) = 0$ for all $i \geq 2$. However, we can write
\begin{equation*}
\begin{split}
\frac{\partial \mathcal{L}}{\partial r_{i+1}}(r^*) - \frac{\partial \mathcal{L}}{\partial r_{i}}(r^*) &= 2(h'(-r_i^*) - h'(-r_{i+1}^*)) - h'(r_i^* - r_{i+1}^*) - h'(r_i^* - r_{i+1}^*)\\
&\quad + \sum_{k=2}^{i-1} (h'(r_k^*-r_i^*) - h'(r_k^* - r_{i+1}^*)) + \sum_{k=i+2}^m (h'(r_{i+1}^*-r_k^*) - h'(r_i^*-r_k^*))\\
&\leq -2h'(r_i^* - r_{i+1}^*) < 0
\end{split}
\end{equation*}
where in the first inequality we used the fact that $r_i^* \geq r_{i+1}^*$ and $h'$ is nondecreasing. In the second inequality we used $h'(r_i^* - r_{i+1}^*) \geq h'(0) > 0$. Since the partial derivatives are zero, this is a contradiction. So, it must be that $r_i^* < r_{i+1}^*$ for all $i \geq 2$.

It remains to show that $r_1^* < r_2^*$, i.e., $r_2^* > 0$. Assume towards a contradiction that $r_2^* \leq 0$. We can write
\begin{equation*}
\begin{split}
\sum_{i=2}^m \frac{\partial \mathcal{L}}{\partial r_i}(r^*) = -2\sum_{i=2}^m h'(-r_i^*) + \sum_{i=2}^{m-1} \sum_{j=i+1}^m (h'(r_i^*-r_j^*) - h'(r_i^*-r_j^*)) = -2\sum_{i=2}^m h'(-r_i^*)
\end{split}
\end{equation*}
since every term $h'(r_i^*-r_j^*)$ appears once with a positive sign and once with a negative sign. Thus, we can write
\begin{equation*}
\begin{split}
\sum_{i=2}^m \frac{\partial \mathcal{L}}{\partial r_i}(r^*) + 2\frac{\partial \mathcal{L}}{\partial r_2}(r^*) &= -2\sum_{i=2}^m h'(-r_i^*) - 4 h'(-r_2^*) + 2\sum_{k=3}^{m} h'(r_2^* - r_k^*)\\
&= -6 h'(-r_2^*) + 2 \sum_{k=3}^m (h'(r_2^* - r_k^*) - h'(-r_k^*))\\
&\leq -6 h'(-r_2^*) < 0
\end{split}
\end{equation*}
where in the first inequality we used the fact that $r_2^* \leq 0$ and $h'$ is nondecreasing. In the second inequality we used $h'(-r_2^*) \geq h'(0) > 0$. Since all the partial derivatives are zero, this is a contradiction. So we also have $r_2^* > 0 = r_1^*$.
\end{proof}

\section{Missing Proofs from \Cref{MLP}}
\MLPuniqueRewards*
\begin{proof}
For each $j \in \{2, \dots, m\}$, we have that both the terms $(1-\alpha) \cdot \ell(r_j)$ and $\alpha \cdot \ell(-r_j)$ appear in $\mathcal{L}$. Since $\ell$ is convex and $\ell'(0) > 0$, it follows that $\lim_{x \to +\infty} \ell(x) = +\infty$. Furthermore, recall that $\ell$ is lower bounded over $\mathbb{R}$. As a result, it follows that $\mathcal{L}(r) \to +\infty$ when $|r_j| \to +\infty$.

Consider the function $\phi_c: (r_2, \dots, r_m) \mapsto \mathcal{L}(c,r_2+c,r_2+c,r_3+c\dots,r_m+c)$ for each $c\in \R$.
It is strictly convex since $\mathcal{L}$ is convex and for every $(r_2, \dots, r_m) \neq (r_2', \dots, r_m')$, there exists $j \in \{2, \dots, m\}$ such that $r_j \neq r_j'$, and thus since the loss function $\ell$ is strictly convex, $\mathcal{L}$ contains a term for which $\ell((r_j+r_j')/2-c) < \ell(r_j-c)/2 + \ell(r_j'-c)/2$.
Thus, together with the previous paragraph, it follows that $\phi_c$ attains its unique minimum $r^*$.
Furthermore, since $\phi_{c_1}=\phi_{c_2}$ for any $c_1,c_2\in \R $, the unique minima of the $\phi_c$ are identical.
It remains to argue that the optimal rewards are ordered $r^*_0 < r_1^* = r_2^* < \dots < r_m^*$. To prove this, we first introduce some additional notation. Define $h: \mathbb{R} \to \mathbb{R}$, $x \mapsto \alpha \cdot \ell(x) + (1-\alpha) \cdot \ell(-x)$. Note that $h$ is convex and differentiable, and thus its derivative $h'$ is non-decreasing. Furthermore, it satisfies
$$h'(0) = \alpha \cdot \ell'(0) - (1-\alpha) \cdot \ell'(0) = (2\alpha - 1) \cdot \ell'(0) > 0$$
since $\alpha > 1/2$ and $\ell'(0)>0$.

We can rewrite the total loss function as
\begin{equation}\label{eq:MLP-total-loss-degenerate}
\begin{split}
 \mathcal{L}(r_0,r_2,r_2,r_3\dots,r_m) & = \alpha \left( \ell(r_2-r_2) +
\sum_{\substack{0 \le i < j \le m \\ i \ne 1 \ \text{or}\ j \ne 2}} \ell(r_i-r_j)
 \right)\\
&+ (1-\alpha) \sum_{0 \leq i < j \leq m} \ell(r_j-r_i)\\& = 
\ell(0) + 2h(r_0-r_2)+ \sum_{j=3}^m h(r_0-r_j)+2\sum_{j=3}^m h(r_2-r_j) + \sum_{\substack{3 \le i < j \le m }} h(r_i - r_j)\end{split}
\end{equation}
with $r_1=r_2$ and thus the partial derivatives for all $i > 2$ as
$$\frac{\partial \mathcal{L}}{\partial r_{i}}(r) = - \sum_{k=0}^{i-1} h'(r_k - r_i) + \sum_{k=i+1}^{m} h'(r_i - r_k),$$
for $i=1,2$
$$\frac{\partial \mathcal{L}}{\partial r_{2}}(r) = - 2 h'(r_0 - r_2) +2\sum_{k=3}^{m} h'(r_2 - r_k),$$
and finally for
$i=0$

$$\frac{\partial \mathcal{L}}{\partial r_{0}}(r) = 2 h'(r_0 - r_2) +\sum_{k=3}^{m} h'(r_0 - r_k).$$

Assume towards a contradiction that $r_i^* \geq r_{i+1}^*$ for some $i \in \{3, \dots, m-1\}$. Since $r^*$ is the optimal solution and $\mathcal{L}$ is differentiable, we have $\partial \mathcal{L}/\partial r_i(r^*) = 0$ for all $i \geq 3$. However, we can write
\begin{equation*}
\begin{split}
\frac{\partial \mathcal{L}}{\partial r_{i+1}^*}(r^*) - \frac{\partial \mathcal{L}}{\partial r_{i}^*}(r^*) &=  \sum_{k=0}^{i-1} h'(r_k^* - r_i^*) - \sum_{k=i+1}^{m} h'(r_i^* - r_k)\\
&- \sum_{k=0}^{i} h'(r_k^* - r_{i+1}^*) + \sum_{k=i+2}^{m} h'(r_{i+1}^* - r_k^*)
\\
&\leq -2h'(r_i^* - r_{i+1}^*) < 0
\end{split}
\end{equation*}
where in the first inequality we used the fact that $r_i^* \geq r_{i+1}^*$ and $h'$ is nondecreasing. In the second inequality we used $h'(r_i^* - r_{i+1}^*) \geq h'(0) > 0$. Since the partial derivatives are zero, this is a contradiction. So, it must be that $r_i^* < r_{i+1}^*$ for all $i \geq 3$.
Next, we show that $r_0^* < r_2^*$. Assume towards a contradiction that $r_2^* \leq r_0^*$. We can write
\begin{equation*}
\begin{split}
\frac{\partial \mathcal{L}}{\partial r_2}(r^*)-2\frac{\partial \mathcal{L}}{\partial r_0}(r^*) &= - 2 h'(r^*_0 - r^*_2) +2\sum_{k=3}^{m} h'(r^*_2 - r^*_k) - 4 h'(r^*_0 - r^*_2) - 2\sum_{k=3}^{m} h'(r^*_0 - r^*_k) \\&\leq -6 h'(r^*_0-r^*_2) <0
\end{split}
\end{equation*}
where in the first inequality we used the fact that $r_0^* \geq r_{2}^*$ (and thus $r_0^*-r_k^* \geq r_{2}^*-r_k^*$) and that $h'$ is nondecreasing.
In the last inequality we used that $r_2^*\leq r_0^*$ and thus $h'(r_0^*-r_2^*)\geq h'(0) >0$. Since the partial derivatives are $0$, this is a contradiction.
Finally we show that $r_3^*>r_2^*$. Again, towards a contradiction assume that $r_3^*\leq r_2^*$.
\begin{equation*}
\begin{split}
2\frac{\partial \mathcal{L}}{\partial r_3}(r^*)-\frac{\partial \mathcal{L}}{\partial r_2}(r^*) &= - 2\sum_{k=0}^{2} h'(r^*_k - r^*_3) + 2\sum_{k=4}^{m} h'(r^*_3 - r^*_k) + 2h'(r^*_0 - r^*_2) -2\sum_{k=3}^{m} h'(r^*_2 - r^*_k)\\&\leq 
-6 h'(r^*_2-r^*_3)-2h'(r^*_0-r^*_3)+2 h'(r^*_0-r^*_2) \\& \leq -6 h'(r^*_2-r^*_3) <0
\end{split}
\end{equation*}
where in the first inequality we used the fact that $r_2^* \geq r_{3}^*$ (and thus $r_2^*-r_k^* \geq r_{3}^*-r_k^*$) and that $h'$ is non-decreasing. In the second inequality we used the same facts and in particular that $r_0^*-r_2^* \leq r_{0}^*-r_3^*$.
In the last inequality we used that $r_3^*\leq r_2^*$ and thus $h'(r_2^*-r_3^*)\geq h'(0) >0$. Since the partial derivatives are $0$, we once again obtain a contradiction.
\end{proof}

\section{Missing Proofs from \Cref{sec:sampling}}

\ellipsefailsPO*
\begin{proof}
Let \(\mu\) be arc-length measure on \(S^1\), so comparison directions are
uniformly distributed. For \(\eta>0\), let \(E_\eta\) be the centered ellipse
with horizontal semiaxis \(1\) and vertical semiaxis \(1+\eta\). We assign
each direction \(x\) the comparison length \(R_\eta(x)\) given by the
distance from the origin to the ellipse in that direction. Thus, writing
\(x(\varphi)=(\cos\varphi,\sin\varphi)\), we have
\(R_\eta(x(\varphi))=(\cos^2\varphi+\sin^2\varphi/(1+\eta)^2)^{-1/2}\).
In particular, \(R_\eta\) is positive, smooth, and even.

Let \(v=(1,1)/\sqrt2\), and let \(Q_t\) denote counterclockwise rotation by
angle \(t\). We consider two voters \(v_1=v\) and
\(v_2(\delta)=-Q_{-\delta^2}v\), with weights
\(p_\delta=1/2+\delta\) and \(1-p_\delta=1/2-\delta\), respectively, where
\(0<\delta<1/2\). Thus, the voter directions are nearly opposite, with a
slightly larger weight for \(v_1\). The rotation by \(\delta^2\) creates
directions preferred by both voters, while affecting the moment vector
only at a smaller order than the weight imbalance \(\delta\).
We will first choose \(\eta>0\) and then \(\delta>0\),
both sufficiently small, so that the optimal reward violates PO.

Let \(q_\delta(x)\) denote the total weight of voters who prefer direction
\(x\), and let \(\ell(t)=\log(1+e^t)\). We consider the following objective:
\begin{equation}\label{eq:ellipse-BT-loss}
\mathcal L_\delta(\theta)
:=\int_{S^1}\left[
q_\delta(x)\ell\bigl(-R_\eta(x)\langle\theta,x\rangle\bigr)
+(1-q_\delta(x))\ell\bigl(R_\eta(x)\langle\theta,x\rangle\bigr)
\right]d\mu(x).
\end{equation}
Since \(R_\eta\) is even, this is twice the weighted version of
\eqref{eq:loss-general-directions}, and hence has the same minimizers.
We define the moment vector and matrix
\[
m_\eta(u):=\int_{\langle u,x\rangle\geq0}R_\eta(x)x\,d\mu(x),
\qquad
S_\eta:=\int_{S^1}R_\eta(x)^2xx^\top\,d\mu(x).
\]
The evenness of \(R_\eta\) implies \(m_\eta(-u)=-m_\eta(u)\), and
\(S_\eta\) is positive definite. We first show how the optimal reward
behaves when the voter directions are nearly opposite and their weights
are nearly equal.

\begin{claim}\label{clm:ellipse-minimizer}
For each fixed \(\eta>0\) and all sufficiently small \(\delta>0\), the
objective \eqref{eq:ellipse-BT-loss} has a unique minimizer
\(\theta_\delta\), satisfying
\[
\theta_\delta=8\delta S_\eta^{-1}m_\eta(v)+O(\delta^2).
\]
\end{claim}

\begin{proof}
First, note that the objective is strictly convex, since the BT loss is
strictly convex and any two distinct parameters have different margins
in all but two directions
on \(S^1\). Now let \(\sigma(t)=(1+e^{-t})^{-1}\), and define
\[
F_\eta(\theta):=\int_{S^1}R_\eta(x)x
\left(\sigma\bigl(R_\eta(x)\langle\theta,x\rangle\bigr)-\frac12\right)
d\mu(x).
\]
Since \(\ell'(s)=\sigma(s)\) and \(\sigma(-s)=1-\sigma(s)\), the
derivative of the scalar loss \(q\ell(-s)+(1-q)\ell(s)\) is
\(-q\sigma(-s)+(1-q)\sigma(s)=\sigma(s)-q\).
Applying the chain rule to \eqref{eq:ellipse-BT-loss} therefore gives
\[
\begin{aligned}
\nabla\mathcal L_\delta(\theta)
&=\int_{S^1}R_\eta(x)x
\left[\sigma\bigl(R_\eta(x)\langle\theta,x\rangle\bigr)-q_\delta(x)\right]d\mu(x)\\
&=F_\eta(\theta)-b_\delta,
\end{aligned}
\]
where
\(b_\delta:=p_\delta m_\eta(v)+(1-p_\delta)m_\eta(v_2(\delta))\).
For the second equality, we used
\(\int_{S^1}R_\eta(x)x\,d\mu(x)=0\) and
\(\int_{S^1}R_\eta(x)xq_\delta(x)\,d\mu(x)=b_\delta\).
Thus a parameter is stationary, meaning that its gradient is zero,
if and only if \(F_\eta(\theta)=b_\delta\).

We next estimate \(b_\delta\). The positive hemispheres of \(v\)
and \(Q_{-\delta^2}v\) differ on arcs
of total length \(O(\delta^2)\). Since \(R_\eta\) is bounded, this gives
\(m_\eta(v_2(\delta))=-m_\eta(v)+O(\delta^2)\). Substituting the voter
weights yields \(b_\delta=2\delta m_\eta(v)+O(\delta^2)\).

We will use the inverse function theorem to solve \(F_\eta(\theta)=b_\delta\) near zero.
For a function
\(F:\mathbb R^m\to\mathbb R^m\), the derivative \(DF(a)\) is the
\emph{Jacobian matrix}: its \((i,j)\)-entry is
\(\frac{\partial F_i}{\partial x_j}(a)\), the rate at which output
coordinate \(i\) changes when input coordinate \(j\) changes.
This matrix describes the linear approximation
\(F(a+h)\approx F(a)+DF(a)h\) for small \(h\).

\begin{theorem}[Inverse Function Theorem {\citep[Theorem~C.34]{lee2013smooth}}]\label{thm:inverse-function}
Let \(U\subseteq\mathbb R^m\) be open, let \(F:U\to\mathbb R^m\) be
smooth,\footnote{A map is smooth, or \(C^\infty\), if all of its partial
derivatives of every order exist and are continuous.} and suppose that
\(DF(a)\) is invertible at \(a\in U\). Then there are open neighborhoods
\(V\subseteq U\) of \(a\) and \(W\) of \(F(a)\) such that, for every
\(b\in W\), the equation \(F(x)=b\) has exactly one solution \(x\in V\).
The map \(G:W\to V\) assigning this solution to \(b\) is smooth and
satisfies \(DG(F(a))=[DF(a)]^{-1}\).
\end{theorem}

To apply this theorem, we first check that \(DF_\eta(0)\) is invertible.
The map \(F_\eta\) is smooth, with \(F_\eta(0)=0\). Since
\(\sigma'(0)=1/4\), the chain rule gives
\[
DF_\eta(0)
=\sigma'(0)\int_{S^1}R_\eta(x)^2xx^\top\,d\mu(x)
=\frac14S_\eta.
\]
For every nonzero \(z\in\mathbb R^2\), we have
\[
z^\top S_\eta z
=\int_{S^1}R_\eta(x)^2\langle z,x\rangle^2\,d\mu(x)>0,
\]
because \(R_\eta(x)>0\) and \(\langle z,x\rangle\neq0\) except at the
two directions perpendicular to \(z\), which have measure zero.
Thus \(S_\eta\) is positive definite, so \(DF_\eta(0)\) is invertible.
By \Cref{thm:inverse-function}, \(F_\eta\) has a smooth local inverse at zero, whose
derivative is \(4S_\eta^{-1}\). Since \(b_\delta\to0\) as \(\delta\to0\)
for fixed \(\eta\), the local inverse applies for all sufficiently small
\(\delta>0\). Taylor expansion of this inverse gives a stationary point
\(\theta_\delta=4S_\eta^{-1}b_\delta+O(\|b_\delta\|^2)\).
By strict convexity, this is the unique global minimizer.
Substituting the estimate for \(b_\delta\) yields the claimed expansion.
\end{proof}

It remains to show that, for a suitable ellipse, the minimizer lies
outside the cone generated by the voters.

\begin{claim}\label{clm:ellipse-moments}
For all sufficiently small \(\eta>0\),
\(\det(v,S_\eta^{-1}m_\eta(v))<0\).
\end{claim}

\begin{proof}
We first expand the radial function as
\(R_\eta(x(\varphi))=1+\eta\sin^2\varphi+O(\eta^2)\), uniformly in
\(\varphi\). The directions preferred by \(v=(1,1)/\sqrt2\) have angles
in \([-\pi/4,3\pi/4]\). Since \(x(\varphi)=(\cos\varphi,\sin\varphi)\),
the two coordinates of \(m_\eta(v)\) are obtained by integrating
\(R_\eta(x(\varphi))\cos\varphi\) and \(R_\eta(x(\varphi))\sin\varphi\),
respectively. In particular, the first (horizontal) coordinate, denoted
\(\bigl(m_\eta(v)\bigr)_1\), satisfies
\[
\begin{aligned}
\bigl(m_\eta(v)\bigr)_1
&=\int_{-\pi/4}^{3\pi/4}(1+\eta\sin^2\varphi)\cos\varphi\,d\varphi+O(\eta^2)\\
&=\left[\sin\varphi+\frac{\eta}{3}\sin^3\varphi\right]_{-\pi/4}^{3\pi/4}+O(\eta^2)\\
&=\sqrt2\left(1+\frac{\eta}{6}\right)+O(\eta^2).
\end{aligned}
\]
The second entry of \(m_\eta(v)\) is obtained similarly. For \(S_\eta\),
the diagonal entries are the integrals of \(R_\eta(x(\varphi))^2\cos^2\varphi\)
and \(R_\eta(x(\varphi))^2\sin^2\varphi\) over \([0,2\pi]\), while the
off-diagonal entries vanish by reflection symmetry. Thus,
\[
\begin{aligned}
m_\eta(v)&=\sqrt2\left(1+\frac{\eta}{6},\,1+\frac{5\eta}{6}\right)
+O(\eta^2),\\
S_\eta&=\pi\begin{pmatrix}1+\eta/2&0\\0&1+3\eta/2\end{pmatrix}
+O(\eta^2).
\end{aligned}
\]
For the second expansion, we use
\(R_\eta(x(\varphi))^2=1+2\eta\sin^2\varphi+O(\eta^2)\).
Combining these expansions gives
\(S_\eta^{-1}m_\eta(v)=\frac{\sqrt2}{\pi}(1-\eta/3,\,1-2\eta/3)
+O(\eta^2)\), and hence
\(\det(v,S_\eta^{-1}m_\eta(v))=-\eta/(3\pi)+O(\eta^2)<0\) for
sufficiently small \(\eta>0\).
\end{proof}

Finally, fix \(\eta>0\) as in \Cref{clm:ellipse-moments}. By
\Cref{clm:ellipse-minimizer}, we can then choose \(\delta>0\) sufficiently
small that \(\det(v,\theta_\delta)<0\). Both voters lie on or
counterclockwise from the ray through \(v\), whereas \(\theta_\delta\)
lies clockwise from it. To obtain a strict PO violation, let
\(u=Q_{\pi/2}v\). Then \(\langle v,u\rangle=0\),
\(\langle v_2(\delta),u\rangle=\sin(\delta^2)>0\), and
\(\langle\theta_\delta,u\rangle=\det(v,\theta_\delta)<0\).
For sufficiently small \(t>0\), the unit vector
\(x=(u+tv)/\|u+tv\|\) therefore satisfies
\(\langle v_1,x\rangle>0\), \(\langle v_2(\delta),x\rangle>0\), and
\(\langle\theta_\delta,x\rangle<0\). Since \(R_\eta(x)>0\), these signs
also hold for the comparison vector \(R_\eta(x)x\). Consider two alternatives
with embeddings \(a=R_\eta(x)x\) and \(b=0\). The inequalities above imply
that both voters prefer \(a\) to \(b\), whereas the optimal reward ranks
\(b\) above \(a\). This violates PO and completes the proof.
\end{proof}

\limitcloserbetter*
\begin{proof}
First of all, note that since $\|\theta\|_2 = \|\theta'\|_2$ and $\inner{\theta'}{v_i} > \inner{\theta}{v_i}$, it must be that $\theta \neq 0$ and $\theta' \neq 0$. Without loss of generality, since the statement of the claim does not change if we rotate the space, we can assume that $v_i = e_1$ is the unit length vector with entry $1$ in the first dimension and $0$ otherwise.

Next, we argue that, without loss of generality, we can assume that $\theta_2 \geq 0$ and $\theta_j = 0$ for all $j \geq 3$, and similarly for $\theta'$. Let $R': \mathbb{R}^{d-1} \to \mathbb{R}^{d-1}$ denote the rotation around the origin in $(d-1)$-dimensional space that maps $(\theta_2, \dots, \theta_d)$ to $(\alpha, 0, \dots, 0)$, where $\alpha \geq 0$. Define $R: \mathbb{R}^d \to \mathbb{R}^d$ by $R(x) = (x_1,R'(x_2,\dots,x_d))$. Note that $\|R(\theta)\|_2 = \|\theta\|_2$ and $\inner{R(\theta)}{v_i} = \inner{\theta}{v_i}$, since $v_i=e_1$. We will show that $\totalloss_i(R(\theta)) = \totalloss_i(\theta)$. Note that $R$ is a bijection and we have
\begin{equation*}
\begin{split}
\inner{R(\theta)}{x} = \theta_1 x_1 + \inner{R'(\theta_2, \dots, \theta_d)}{(x_2,\dots, x_d)} &= \theta_1 x_1 + \inner{(\theta_2, \dots, \theta_d)}{R'^{-1}(x_2,\dots, x_d)}\\
&= \inner{\theta}{R^{-1}(x)}
\end{split}
\end{equation*}
where $R'^{-1}$ is the inverse rotation to $R'$. We can thus write
\begin{equation*}
\begin{split}
\totalloss_i(R(\theta)) = \int_{x \in S^{d-1}, x_1 \geq 0} \ell(-\inner{R(\theta)}{x}) \, dx &= \int_{x \in S^{d-1}, x_1 \geq 0} \ell(-\inner{\theta}{R^{-1}(x)}) \, dx\\
&= \int_{x \in S^{d-1}, x_1 \geq 0} \ell(-\inner{\theta}{x}) \, dx\\
&= \totalloss_i(\theta)
\end{split}
\end{equation*}
since $R^{-1}$ is a smooth bijection that preserves distances. Thus, we can assume that $\theta_2 \geq 0$ and $\theta_j = 0$ for all $j \geq 3$, and similarly for $\theta'$.

Let $T$ be the rotation that maps $\theta$ to $\theta'$. We can write
\begin{equation*}
\begin{split}
\totalloss(\theta) &= \int_{x \in S^{d-1}, x_1 \geq 0} \ell(-\inner{\theta}{x}) \, dx\\
&= \int_{x \in S^{d-1}, [T^{-1}(x)]_1 \geq 0} \ell(-\inner{\theta}{T^{-1}(x)}) \, dx\\
&= \int_{x \in S^{d-1}, [T^{-1}(x)]_1 \geq 0} \ell(-\inner{T(\theta)}{x}) \, dx\\
&= \int_{x \in S^{d-1}, [T^{-1}(x)]_1 \geq 0, x_1 \geq 0} \ell(-\inner{\theta'}{x}) \, dx + \int_{x \in S^{d-1}, [T^{-1}(x)]_1 \geq 0, x_1 \leq 0} \ell(-\inner{\theta'}{x}) \, dx.
\end{split}
\end{equation*}
We can also decompose
\begin{equation*}
\begin{split}
\totalloss(\theta') &= \int_{x \in S^{d-1}, x_1 \geq 0} \ell(-\inner{\theta'}{x}) \, dx\\
&= \int_{x \in S^{d-1}, [T^{-1}(x)]_1 \geq 0, x_1 \geq 0} \ell(-\inner{\theta'}{x}) \, dx + \int_{x \in S^{d-1}, [T^{-1}(x)]_1 \leq 0, x_1 \geq 0} \ell(-\inner{\theta'}{x}) \, dx.
\end{split}
\end{equation*}
Thus, we obtain
\begin{equation*}
\begin{split}
\totalloss(\theta) - \totalloss(\theta') &= \int_{x \in S^{d-1}, [T^{-1}(x)]_1 \geq 0, x_1 \leq 0} \ell(-\inner{\theta'}{x}) \, dx - \int_{x \in S^{d-1}, [T^{-1}(x)]_1 \leq 0, x_1 \geq 0} \ell(-\inner{\theta'}{x}) \, dx\\
&= \int_{x \in S^{d-1}, [T^{-1}(x)]_1 \leq 0, x_1 \geq 0} \ell(\inner{\theta'}{x}) - \ell(-\inner{\theta'}{x}) \, dx\\
&= \int_{x \in S^{d-1}, [T^{-1}(x)]_1 < 0, x_1 > 0} \ell(\inner{\theta'}{x}) - \ell(-\inner{\theta'}{x}) \, dx.
\end{split}
\end{equation*}
Note that the domain $\{x \in S^{d-1}, [T^{-1}(x)]_1 < 0, x_1 > 0\}$ has positive measure, since $\inner{\theta'}{v_i} > \inner{\theta}{v_i}$. Furthermore, we have $\ell(t) > \ell(-t)$ for all $t > 0$.\footnote{This follows from the fact that $\ell$ is nondecreasing and $\ell'(0) > 0$.} As a result, if we can show that $\inner{\theta'}{x} > 0$ for all $x \in S^{d-1}$ with $[T^{-1}(x)]_1 < 0$ and $x_1 > 0$, then we will obtain $\totalloss(\theta) - \totalloss(\theta') > 0$, as desired.

Recall that we assume without loss of generality that $\theta_2, \theta_2' \geq 0$ and $\theta_j = \theta_j' = 0$ for all $j \geq 3$. Furthermore, by assumption we have $\theta_1' > \theta_1$. Consider any $x \in S^{d-1}$ with $\inner{\theta'}{x} \leq 0$ and $x_1 > 0$. Assume towards a contradiction that $[T^{-1}(x)]_1 < 0$. The rotation $T^{-1}$ maps $\theta'$ to $\theta$, which both lie in the upper hemisphere, and $\theta_1' > \theta_1$. Thus, the rotation is counter-clockwise and of angle at most $\pi$. Since $x_1 > 0$, we can assume without loss of generality that $\theta_1 < 0$ and $\theta_2=0$, i.e., $\theta$ is in the negative $x$ axis direction. Indeed, since $x_1 > 0$, if rotating $x$ in the counterclockwise direction by some angle $\alpha < \pi$ yields a point $y$ with $y_1 < 0$, then rotating it by any angle $\beta \in (\alpha, \pi]$ will also yield a point $y$ with $y_1 < 0$. Now it is easy to see that the rotation $T^{-1}$ that maps $\theta'$ to the negative axis direction also maps any point $x$ with $\inner{\theta'}{x} \leq 0$ to a point $y = T^{-1}(x)$ with $y_1 \geq 0$, a contradiction.
\end{proof}

\limitminattained*

\begin{proof}
We show that for any $M\in \mathbb{R}$, there exists a $t>0$ such that if $\|\theta\|_2 > t$ then $\mathcal{L}(\theta)>M$. Then, since $\mathcal{L}(0)$ is finite, we can conclude that $\mathcal{L}$ attains its minimum. Observe that since two voters, say $v_1$ and $v_2$, have distinct directions, there exists $\kappa<1$ such that for any $\theta \in S^{d-1}$, there exists $i \in \{1,2\}$ such that $\inner{\theta}{v_i} \leq \kappa$. Here we assume that $\|v_1\|_2 = \|v_2\|_2 = 1$ without loss of generality.

Let $M$ be arbitrary. Let $\theta \in S^{d-1}$ be arbitrary. Without loss of generality, say that voter $v_1= e_1$ is such that $\inner{\theta}{v_1} \leq \kappa$. Since the loss function $\ell$ is lower bounded, there exists $K < 0$ such that
\begin{equation*}
\begin{split}
\totalloss(t\theta) \geq K + \totalloss_1(t\theta) &\geq K + \int_{x \in S^{d-1}, x_1 \geq 0} \ell(-\inner{t\theta}{x}) \, dx\\
&\geq 2K + \int_{x \in S^{d-1}, x_1 \geq 0, \inner{\theta}{x} \leq f(\kappa, d)} \ell(-\inner{t\theta}{x}) \, dx
\end{split}
\end{equation*}
where $f(\kappa, d) < 0$ is such that the set $\{x \in S^{d-1}: x_1 \geq 0, \inner{\theta}{x} \leq f(\kappa, d)\}$ has strictly positive measure bounded away from zero by at least $g(\kappa, d) > 0$. As a result, we obtain that
$$\totalloss(t\theta) \geq -2K + g(\kappa,d) \cdot \ell(-t \cdot f(\kappa,d)) > M$$
for a sufficiently large $t > 0$, since $\ell$ is strictly increasing for positive inputs.\footnote{This follows from the fact that $\ell$ is convex and $\ell'(0) > 0$.}
\end{proof}

\limitnotzero*
\begin{proof}
Let $z \in S^{d-1}$ be such that $\inner{v_i}{z} > 0$ for all voters $i \in [n]$. We can write
$$\totalloss_i(t \cdot z) = \int_{x \in S^{d-1}, \inner{v_i}{x} \geq 0} \ell(-\inner{t \cdot z}{x}) \, dx.$$
Taking the derivative with respect to $t$ at $t=0$ we obtain
$$\frac{d}{dt}\big|_{t=0}\totalloss_i(t \cdot z) = \int_{x \in S^{d-1}, \inner{v_i}{x} \geq 0} -\inner{z}{x} \ell'(0) \, dx = \ell'(0) \int_{x \in S^{d-1}, \inner{v_i}{x} \geq 0} -\inner{z}{x} \, dx < 0$$
which follows from $\ell'(0) > 0$ and
$$\int_{x \in S^{d-1}, \inner{v_i}{x} \geq 0} \inner{z}{x} \, dx > 0$$
which follows from $\inner{v_i}{z} > 0$ using similar ideas to those in the proof of \Cref{clm:limit-closer-better}.
\end{proof}

\noPMC*

\begin{proof}
Consider $\mathbb{R}^2$ and let $v_1 = (0,1)$, $v_2 = (1,0)$. For $p \in [1/4,1/2]$, we consider the instance where a fraction $p$ of the voters are in direction $v_1$, and the remaining fraction $(1-p)$ are in direction $v_2$. Thus, the total loss function that is minimized over $\theta \in \mathbb{R}^2$ can equivalently be written as
\[
\totalloss(\theta,p) := 2p \int_{x \in S^{1}, x_2 \geq 0} \ell(-\inner{\theta}{x}) \, dx + 2(1-p)\int_{x \in S^{1}, x_1 \geq 0} \ell(-\inner{\theta}{x}) \, dx.
\]

By \Cref{clm:limit-min-attained}, $\totalloss(\cdot, p)$ attains its minimum for each $p \in [1/4,1/2]$. Furthermore, since $\ell$ is strictly convex, so is\footnote{This follows from standard results in convex analysis~\citep{Rockafellar1970-convex-analysis} together with the observation that for any $a \neq b$, the set of points $x$ on $S^1$ such that $\inner{a - b}{x} = 0$ has measure zero (on $S^1$).} $\totalloss(\cdot,p)$, and thus the minimizer of $\totalloss(\cdot,p)$ is unique.

For any $p < 1/2$, $v_2$ is the strict majority voter and so $\theta = v_2$ is the only pairwise majority consistent direction. In order to show that PMC is not satisfied, it thus suffices to prove that for some $p < 1/2$, the minimizer of the total loss is different from direction $v_2$.

First, for $p=1/2$ we can write
\begin{align*}
\totalloss(\theta,1/2)
&= \int_{x \in S^{1}, x_2 \geq 0} \ell(-\inner{\theta}{x}) \, dx
 + \int_{x \in S^{1}, x_1 \geq 0} \ell(-\inner{\theta}{x}) \, dx \\
&= \int_{x \in S^{1}, x_1 \geq 0} \bigl(\ell(-\inner{R(\theta)}{x})+\ell(-\inner{\theta}{x}) \bigr) \, dx,
\end{align*}
where $R$ denotes the $90^\circ$ clockwise rotation, i.e., $R(\theta) = (\theta_2,-\theta_1)$.
Let $\theta' = (\theta_2,\theta_1)$ and note that $\inner{R(\theta)}{x} = \theta_2 x_1 - \theta_1 x_2 = \inner{\theta'}{(x_1,-x_2)}$. Since the domain of integration is symmetric with respect to the transformation $(x_1,x_2) \mapsto (x_1,-x_2)$, we obtain
\[
\totalloss(\theta,1/2)
= \int_{x \in S^{1}, x_1 \geq 0} \bigl(\ell(-\inner{\theta'}{x}) + \ell(-\inner{\theta}{x})\bigr)\, dx.
\]
It follows that $\totalloss(\theta,1/2) = \totalloss(\theta',1/2)$, i.e., the loss is invariant under permuting $\theta_1$ and $\theta_2$. By convexity,
\[
\totalloss\bigl((\theta+\theta')/2,1/2\bigr)
\leq \frac{1}{2}\bigl(\totalloss(\theta,1/2) + 
\totalloss(\theta',1/2)\bigr)=\totalloss(\theta,1/2),
\]
so the unique minimizer $\theta^*$ of $\totalloss(\cdot,1/2)$ satisfies $\theta_1^* = \theta_2^*$. From \Cref{clm:limit-not-zero}, we know $\theta^* \neq 0$. By \Cref{thm:limit-PO}, PO holds for this instance, so $\theta^*$ lies in the positive cone of $v_1$ and $v_2$, and in particular $\theta_2^* > 0$.

Now let $M > 3\|\theta^*\|$ and $B(M) = \{\theta \in \mathbb{R}^2 : \|\theta\| \leq M\}$. The map $\totalloss \colon  B(M) \times [1/4,1/2] \to \mathbb{R}$ is continuous, so by Berge's theorem (\Cref{thm:berge}) the function\footnote{For each $p \in [1/4,1/2]$, the minimizer is unique.} 
\[
\totalloss^* \colon [1/4,1/2] \to B(M),\qquad
\totalloss^*(p)=\argmin_{\theta \in B(M)} \totalloss(\theta,p)
\]
is continuous. Thus, for sufficiently small $\delta \in (0,1/4)$,
\[
\|\totalloss^*(1/2-\delta) - \theta^*\| < \theta_2^* \leq \|\theta^*\|,
\]
so
\[
\|\totalloss^*(1/2-\delta)\|
\leq \|\theta^*\| + \|\totalloss^*(1/2-\delta) - \theta^*\|
\leq 2\|\theta^*\| < M.
\]
In particular, $\totalloss^*(1/2-\delta)$ does not lie on the boundary of $B(M)$, and therefore also minimizes $\totalloss(1/2-\delta,\theta)$ over all $\theta \in \mathbb{R}^2$.

Finally, any $(t,0)$ with $t>0$ has distance at least $\theta_2^*$ from $\theta^*$, so $\totalloss^*(1/2-\delta)$ is not a scalar multiple of the majority voter $v_2 = (1,0)$, and therefore PMC is not satisfied when $p = 1/2 - \delta$.
\end{proof}

\end{document}